\documentclass[preprint2]{aastex}
\let\captionbox\relax
\usepackage[latin1]{inputenc}
\usepackage{amsmath}
\usepackage{wasysym}
\usepackage{amsfonts}
\usepackage{amssymb}
\usepackage{graphicx}
\usepackage{caption}
\usepackage{subcaption}
\usepackage{minipage}
\usepackage{listings}
\usepackage{natbib}
\shorttitle{Sources of F$_{10.7}$ Flux}
\shortauthors{Schonfeld et al.}

\begin{document}

\title{Coronal Sources of the Solar F$_{10.7}$ Radio Flux}

\author{S. J. Schonfeld$^1$, S. M. White$^2$, C. J. Henney$^2$, C.  N. Arge$^2$, \and R. T. J. McAteer$^1$}
\affil{$^1$New Mexico State University, Department of Astronomy, P.O. Box 30001, MSC 4500, Las Cruces, NM 88003-8001, schonfsj@gmail.com}
\affil{$^2$Air Force Research Laboratory, Space Vehicles Directorate, AFRL, 3550 Aberdeen Ave SE, Kirtland AFB, Albuquerque, NM 87117}

\begin{abstract}
We present results from the first solar full-disk F$_{10.7}$ (the radio flux at $10.7$ cm, $2.8$ GHz) image taken with the S-band receivers on the recently upgraded Karl G. Jansky Very Large Array (VLA) in order to assess the relationship between the F$_{10.7}$ index and solar extreme ultra-violet (EUV) emission. To identify the sources of the observed $2.8$ GHz emission, we calculate differential emission measures (DEMs) from EUV images collected by the Atmospheric Imaging Assembly (AIA) and use them to predict the bremsstrahlung component of the radio emission. By comparing the bremsstrahlung prediction and radio observation we find that $8.1\pm 0.5\%$ of the variable component of the F$_{10.7}$ flux is associated with the gyroresonance emission mechanism. Additionally, we identify optical depth effects on the radio limb which may complicate the use of F$_{10.7}$ time series as an EUV proxy. Our analysis is consistent with a coronal iron abundance that is $4$ times the photospheric level.

\end{abstract}


\maketitle

\section{Introduction}
\label{sec:introduction}
Microwave emission from the sun, specifically the F$_{10.7}$ index (the solar radio flux at 10.7 cm, 2.8 GHz), has long been known to correlate with solar extreme ultra-violet (EUV) emission on time-scales of days and longer \citep{Covington1948, Covington1951, Covington1955, Vats1998, Foukal1998, Tapping2003a} and has been measured daily since 1947 \citep{Covington1969, Tapping1987}. This correlation and the transparency of the atmosphere to microwave signals \citep{Tapping2013} has led to the use of F$_{10.7}$ as a proxy measurement for solar EUV irradiance, which heats and ionizes the Earth's atmosphere but cannot be observed from the ground. The observed F$_{10.7}$ signal was one of the inputs for the original terrestrial ionospheric and thermospheric modelling efforts \citep{Bhatnagar1966, Jacchia1971} and remains one of the primary model inputs today, even with the availability of direct EUV observations \citep{Tobiska2008}. F$_{10.7}$ is often preferred over other proxies of solar activity such as Sunspot Number (SSN), the MgII core-to-wing index, Ly$\alpha$ irradiance, etc. due to its high degree of correlation with solar EUV output which results from the fact that the main variable components of both originate in the same coronal plasma \citep{Swarup1963}.

As early as \cite{Kundu1959} it was understood that thermal microwave sources could be split into three classes: a low intensity background originating from the quiet Sun \citep{Martyn1948}, a moderate intensity signal seen in and around plage and active regions \citep{Covington1947}, and a high intensity component commonly associated with active region cores \citep{Piddington1951}. The background component (here taken to be $65.2\pm 2.0$ sfu of the F$_{10.7}$ flux) can be explained with a uniform optically thick chromosphere of $11,000$ K and an overlying optically thin $10^{6}$ K corona \citep{Zirin1991}, which is generally removed for the purposes of EUV approximation \citep[but see also ][]{Landi2003,Landi2008}. Numerous studies have investigated the variability of long duration coronal microwave (primarily F$_{10.7}$) timeseries for use as proxies of EUV \citep{Covington1951, Covington1969, Tapping1987,Wilson1987, Lean1989, Bouwer1992, Foukal1998, Parker1998, Vats1998, Tobiska2008, DudokdeWit2009, Maruyama2010, Svalgaard2010a, Chen2011a, Johnson2011, Maruyama2011, Henney2012a, Deng2013a} and total solar irradiance \citep{Tapping2007, Frohlich2009, Tapping2011} but there is still considerable debate as to the exact source of the time--variable component. Some studies have argued that the variable microwave component is optically thin bremsstrahlung emission originating from the plage regions \citep{Felli1981, Tapping1990, Tapping2003a} while others suggested that it is primarily gyroresonance emission from the strong magnetic fields in active region cores \citep{Schmahl1995, Schmahl1998}. As discussed below (in \S \ref{sec:physical_background}), bremsstrahlung emission is closely related to EUV emission, while gyroresonance emission is not. A recent study by \cite{DudokdeWit2014a} suggests that the gyroresonance component can ``account for $90\%$ of the rotational variability in the F$_{10.7}$ index."

This ambiguity in the source of the signal variability is best resolved through imaging when the individual sources are resolved. Many such studies have been performed throughout the microwave regime, both of the entire solar disk \citep{Swarup1963, Bastian1988a, Gopalswamy1991, Tapping2003a} and of individual active regions \citep{Felli1981,White1992}. There has been less work specifically at $2.8$ GHz, and the best imaging to date was that of \cite{Saint-Hilaire2012} who used the Allen Telescope Array to observe the full Sun between $1.43$ and $6$ GHz (including F$_{10.7}$). They used the emission spectra and polarization to identify gyroresonance sources with a spatial resolution at $2.8$ GHz of about $1\arcmin$. However, due to a lack of available instrumentation, imaging of F$_{10.7}$ with resolution better than one arcminute did not become possible until the upgrade to the Karl G. Jansky Very Large Array (VLA). \cite{Selhorst2014} recently completed a study of the statistical properties of spatially resolved active regions observed at $17$ GHz in order to identify bremsstrahlung and gyroresonance emission and noted that a similar analysis is needed to fully understand F$_{10.7}$. While high-resolution radio studies are necessary to identify gyroresonance regions and their contribution to F$_{10.7}$, they are not in general sufficient to determine the magnitude of the gyroresonance emission \citep{White1997} and some independent estimate of one of the components of the radio signal is required.

The bremsstrahlung component of the flux can be extrapolated without direct radio observation if the differential emission measure (DEM, i.e., the distribution of plasma density with temperature in the corona) is known. DEMs can be calculated from sets of optically thin observations, including EUV imaging and spectroscopy, as long as they are sensitive to a range of coronal temperatures wide enough to sample the dominant coronal plasma. \cite{Landi2003, Landi2008} constructed solar--minimum DEMs using both EUV and radio observations to constrain the plasma structure from the chromosphere, which has significantly fewer EUV emission lines, all the way through the corona. With current instrumentation, coronal DEMs are most commonly constructed from EUV data because of the high spatial, temporal, and spectral resolution of observations, especially in the Solar Dynamics Observatory (SDO) era \citep{Pesnell2011}. Comparisons between radio observations and the predicted bremsstrahlung component based on DEMs computed from EUV images have been made previously, e.g., by \cite{White2000a} and \cite{Zhang2001}. EUV emission is dominated by specific atomic emission lines while the bulk thermal emission in the radio is generated by free electrons and therefore represents the density of fully--ionized hydrogen in the corona. This means that the elemental abundance is a (presumably) constant scaling factor relating the element--specific DEM and the radio measurement. If the abundance is known, then the DEM can be used to predict the optically thin bremsstrahlung emission from the plasma observed in the EUV. \cite{White2000a} used this technique to measure the Fe abundance in an active region.

In this paper we analyze the first full--disk image of F$_{10.7}$ emission acquired with the VLA, which is the highest spatial resolution $2.8$ GHz image to date, and compare it with spatially--resolved EUV images. The prediction of the bremsstrahlung emission along with the polarization signal in the radio are used to identify gyroresonance sources in the radio image and determine the total gyroresonance component of the F$_{10.7}$ flux. In \S \ref{sec:physical_background} the physical connections between the various EUV and radio emission mechanisms are discussed, with details of the observations given in \S \ref{sec:data}. The calculation of the DEM is described in \S \ref{sec:DEM} and the subsequent comparison of the bremsstrahlung prediction and the radio observation is explained in \S \ref{sec:analysis}. \S \ref{sec:discussion} includes a discussion of the results and implications, and we conclude and suggest future additions to this work in \S \ref{sec:conclusion}.

\section{Physical Background}
\label{sec:physical_background}
\subsection{Radio Emission}
There are only two identified mechanisms that produce microwave radio emission from the non-flaring sun: bremsstrahlung and gyroresonance \citep{Kundu1965}.

Thermal bremsstrahlung results from the collisional interaction of electrons and ions \citep{Wild1963}. For optically thin bremsstrahlung emission in coronal conditions the radio flux density (in units of $\text{erg cm}^{-2}\ \text{s}^{-1}\ \text{Hz}^{-1}$) is given by:
\begin{flalign}
f_{\nu}\ =\ & 9.78\times 10^{-3}\frac{2k_{\text{B}}}{c} \left(1+4\frac{N_{\text{He}}}{N_{\text{H}}}\right) \nonumber \\
& \times \int\int T^{-0.5} \text{DEM}(T) G(T)\ \text{d}T\ \text{d}\Omega
\label{eqn:Bremsstrahlung}
\end{flalign}
where $k_{\text{B}}=1.38\times 10^{-16}\ \text{g cm}^{-2}\ \text{s}^{-2}\ \text{K}^{-1}$ is Boltzmann's constant, $c=3\times 10^{10}\ \text{cm s}^{-1}$ is the speed of light, $N_{\text{He}}/N_{\text{H}}=0.085$ \citep{Asplund2009} is the number (or number density) ratio of Helium to Hydrogen in the emitting medium, $T$ is the temperature in Kelvin, $\text{G}(T)=24.5+ln\left(T/\nu\right)$ is the Gaunt factor where $\nu$ is the frequency in Hz, $\text{d}\Omega$ is the solid angle of the source \citep{Dulk1985} and $\text{DEM}(T)$ is the integral along the line of sight through the corona of $\text{d}(n_{e}n_{\text{H}})/\text{d}T$. The dependence of the flux on $n^{2}\,T^{-0.5}$ means that the optically thin bremsstrahlung flux is actually relatively insensitive to the temperature distribution and is much more sensitive to the plasma density (see \S \ref{sec:DEM}).

Thermal bremsstrahlung emission at microwave frequencies ($1-30$ GHz) generally becomes optically thick in the chromosphere because of both the increased density and decreased temperature. The altitude (and therefore temperature and density) at which this optically thick boundary occurs is a strong function of frequency $\nu$ since bremsstrahlung opacity varies as $\nu^{-2}\,n^2\,T^{-0.5}$, with higher frequencies penetrating deeper into the chromosphere. This leads to a frequency dependence in the observed height of the solar limb at microwave frequencies, with the apparent size of the solar disk decreasing with increasing frequency \citep{Furst1979}. In active regions with very high density this optically thick boundary can also occur in the corona at low microwave frequencies, increasing the observed brightness temperature dramatically and blocking observation of the lower atmosphere. It is common for active regions to be optically thick in the corona due to bremsstrahlung at $1.4$ GHz but optically thin at $5$ GHz \citep{White1999}, putting F$_{10.7}$ ($2.8$ GHz) at an interesting transition frequency. Coronal bremsstrahlung emission is generally not strongly polarized, but magnetic fields do break the degeneracy of collisional interactions and produce weak circular polarization \citep{White1997}.

Gyroresonance emission arises from the acceleration of electrons as they spiral around magnetic field lines. At coronal temperatures even thermal electrons have weakly relativistic velocities and produce opacity not just at the gyrofrequency, but also at low order harmonics \citep{Wild1963}. Depending primarily on the magnetic field orientation and the polarization mode, this emission becomes optically thick in the $s=1,\ 2,\ 3,\ \text{or}\ 4$ harmonic of the gyrofrequency:
\begin{equation}
\nu_{B}=2.80B \hspace{0.5cm} \text{[MHz]}
\label{eqn:gyrofrequency}
\end{equation}
where $B$ is the magnitude of the magnetic field in G \citep{White1997}. This means F$_{10.7}$ gyroresonance observations at $2.8$ GHz come from thin surfaces with constant magnetic field strengths of $B=10^3/s=1000,\ 500,\ 333,\ \text{and}\ 250$ G for harmonics $s=1,\ 2,\ 3,\ \text{and}\ 4$ respectively.

The characteristic motion associated with the gyroresonance process naturally causes the emission to be highly circularly polarized (because any intrinsic linear polarization is wiped out by Faraday rotation in the solar atmosphere). Emission with polarization in the sense of an electron spiralling around the field is called the extraordinary or \textit{x}-mode, while polarization with the opposite sense of rotation is called the ordinary or \textit{o}-mode. Electrons couple much more strongly to the \textit{x}-mode than the \textit{o}-mode because of the shared sense of rotation. Consequently, the \textit{x}-mode generally has larger opacity and becomes optically thick in higher (harmonic and altitude) gyroresonant layers. The generally positive temperature gradient in the lower corona means that the \textit{x}-mode then has a higher brightness temperature causing an observed net circular polarization from gyroresonance sources in the sense of the local \textit{x}-mode. 

\subsection{EUV Emission}
While there is high energy X-ray bremsstrahlung continuum emission \citep{Craig1976}, the majority of the energy output from the non-flaring corona comes from collisionally excited atomic emission lines, predominantly observed in the EUV \citep{Golub2010}. At coronal temperatures the EUV spectrum is dominated by emission lines with strengths dependant on a number of characteristics of the bulk plasma and the individual emitting atom. These include but are not limited to: the local plasma density (which influences the collision rate), the local electron temperature (which determines the energy of the collisions), the ionization and excitation state of the atom (which restrict the available transitions), and the oscillator strengths of the available transitions (which determine both the probabilities of each emission as well as their respective energy spectra). However, assuming the atomic details are known, the EUV emission properties of a bulk plasma in ionization equilibrium are completely determined by the relative elemental abundances and the electron density and temperature distribution \citep{Craig1976}. The temperature dependence of individual emission lines means that observing specific regions of the EUV spectrum highlights very different coronal features. One commonality among these coronal EUV observations is that they remain optically thin in all non-flaring conditions, and therefore all coronal plasma above the much denser and optically thick chromosphere is visible.

\subsection{DEM Connection}
As has been noted, both radio bremsstrahlung continuum and EUV emission line strengths are at least partially dependent on the plasma density and temperature distribution which can be described by the DEM, and therefore the two can be compared. We compute the DEM (\S \ref{sec:DEM}) using the higher spatial, temporal, and spectral resolution of EUV observations and then use it to predict the optically thin bremsstrahlung radio emission from the same plasma. To first order, the difference between the optically thin bremsstrahlung prediction and a radio measurement should reflect the optical depth effects in the bremsstrahlung emission and/or the presence of gyroresonance components. Note that the detailed temperature structure of the DEM is less important for this application because the comparison with the radio only depends on the (temperature-weighted) integral over the DEM as shown in equation \ref{eqn:Bremsstrahlung}.

This comparison lacks the information needed to account for the altitude of the optically thick layer in the radio, which adds a complicating factor. Both bremsstrahlung emission from dense active regions and gyroresonance emission can generate optically thick layers in the corona that block radio emission from any plasma below that layer. The EUV lines used to compute the DEM remain optically thin all the way to the chromosphere, therefore the EUV will generally observe more plasma than the radio and may lead to a relative overestimate of the observable optically thin bremsstrahlung emission. This effect is difficult to assess and we discuss it further in \S \ref{sec:uncertainties}.

\section{Data}
\label{sec:data}
The observations for this analysis were taken during the rising phase of solar cycle 24 on 2011 December 9, between 15 and 23 UT. During this period there was moderate solar activity on the earthward hemisphere and the F$_{10.7}$ index was $143.5\pm 1.2$ sfu \citep{Tapping1994}, but there were no recorded solar storms of any kind. The observed variability of coronal features was insignificant and occurred mostly on scales below the resolution of the radio observations.

\subsection{Karl G. Jansky Very Large Array}
At the time of these observations, the Karl G. Jansky Very Large Array (VLA, operated by the National Radio Astronomy Observatory) was in the process of being upgraded, and data were taken as a shared--risk project. As such, only 17 of the VLA's nominal 27 available antennae were equipped with the ``S-band'' feeds ($2-4$ GHz) needed to observe F$_{10.7}$. The VLA was in its most compact (``D'') array configuration, appropriate for recovering the flux of large--scale sources in the solar atmosphere. The reduced number of antennae decreased the resulting image quality significantly, because of both the reduced coverage of the u-v plane, as well as the $37\%$ decrease in collecting area. For this analysis, eight 2 MHz channels centred on $2.783$ GHz were summed for a total bandwidth of $16$ MHz. Solar observations in S-band are taken with the additional nominally-$20$ dB attenuators in the signal path and these add phase changes that are corrected using independently measured values of the delays (Bin Chen, private communication). Unfortunately, the measurements of the primary flux calibration source were corrupted, preventing independent measurement of the amplitude changes due to the attenuators. Consequently, the solar fluxes were calibrated assuming exactly $20$ dB of attenuation and a nominal flux for the secondary calibrator. We estimate that the VLA solar fluxes therefore have an uncertainty of order 20\%.

The field of view of a single VLA pointing with the S-band receivers is nominally $15\arcmin$ and therefore mosaicking is required to image the entire $30\arcmin$ diameter solar disk. A honeycomb pattern mosaic was used, with a single pointing at disk center surrounded by six fields designed to overlap by half a beam width. The center of each field was tracked over the course of the eight hour observation, taking into account solar differential rotation. This led to some slight feature smearing near the edges of each field, but the magnitude was well below the final $25\arcsec$ resolution of the observation and there appeared to be no effect on the final mosaic image. Each field was calibrated separately and then the fields were imaged jointly in a single map using maximum entropy deconvolution. The images were restored with a spatial resolution of $25\arcsec$. Due to the limited field of view, as well as the restriction caused by the minimum baseline, the observations were insensitive to emission on the scale of the solar disk. Attempts to restore this component using a default disk of the right dimension in the deconvolution process were unsuccessful \citep[because they failed to produce a mostly flat disk as seen in the Allen Telescope Array observation of ][]{Saint-Hilaire2012}, and we therefore do not address the spatial distribution of the large--scale emission here. This large--scale component will be analyzed with a subsequent data set acquired using more antennae and a larger mosaic pattern.

\subsection{Nobeyama Radioheliograph}
The $17$ GHz data from the Nobeyama Radioheliograph \citep[NoRH; ][]{Nakajima_abs1994} were used to assess optical depth effects in the bremsstrahlung radio emission (the optical depth of bremsstrahlung at $17$ GHz is $37$ times smaller than at $2.8$ GHz) and as a calibration check. NoRH makes full disk images of the Sun at $17$ and $34$ GHz every day between 23:00 and 06:30 UT. Located in Japan, the dedicated solar array is unable to observe simultaneously with the VLA, and therefore 
the Nobeyama images collected just after the completion of the VLA observation were used.
The data were mapped and calibrated using standard procedures: amplitude calibration assumed that the background disk component that generally dominates the total flux had a brightness temperature of $10^{4}$ K, which is known to be consistent with well--calibrated flux monitoring at this frequency by the Nobeyama polarimeters \citep[NoRP; ][]{Nakajima1985}. The magnetic field dependence of equation \ref{eqn:gyrofrequency} means that $17$ GHz is only sensitive to gyroresonance emission from strong magnetic fields, requiring a coronal field greater than $2000$ G to observe the third harmonic. Coronal magnetic field strengths this high are typically only seen in the case of very large active regions. Given the absence of such regions during the observation, the Nobeyama observations are expected to detect purely bremsstrahlung emission from the solar atmosphere that is excess to the (chromospheric) background brightness temperature level of $10^{4}$ K. The $17$ GHz image was made by synthesizing and deconvolving data taken at $45$ s intervals over a period of several hours, and rotating the final image back in time to match the VLA image. The spatial resolution in the final $17$ GHz image is $12\arcsec$, and the flat background disk of $10^{4}$ K (with a radius 1.0125 times the photospheric radius that fits the $17$ GHz visibilities) is subtracted for the region analysis.

\subsection{Atmospheric Imaging Assembly}
\begin{figure*}[!t]
\centering 
\begin{minipage}{0.49\textwidth}
\centering
\includegraphics[trim=1cm 1cm 1cm 1cm, clip=true, height=0.85\linewidth]{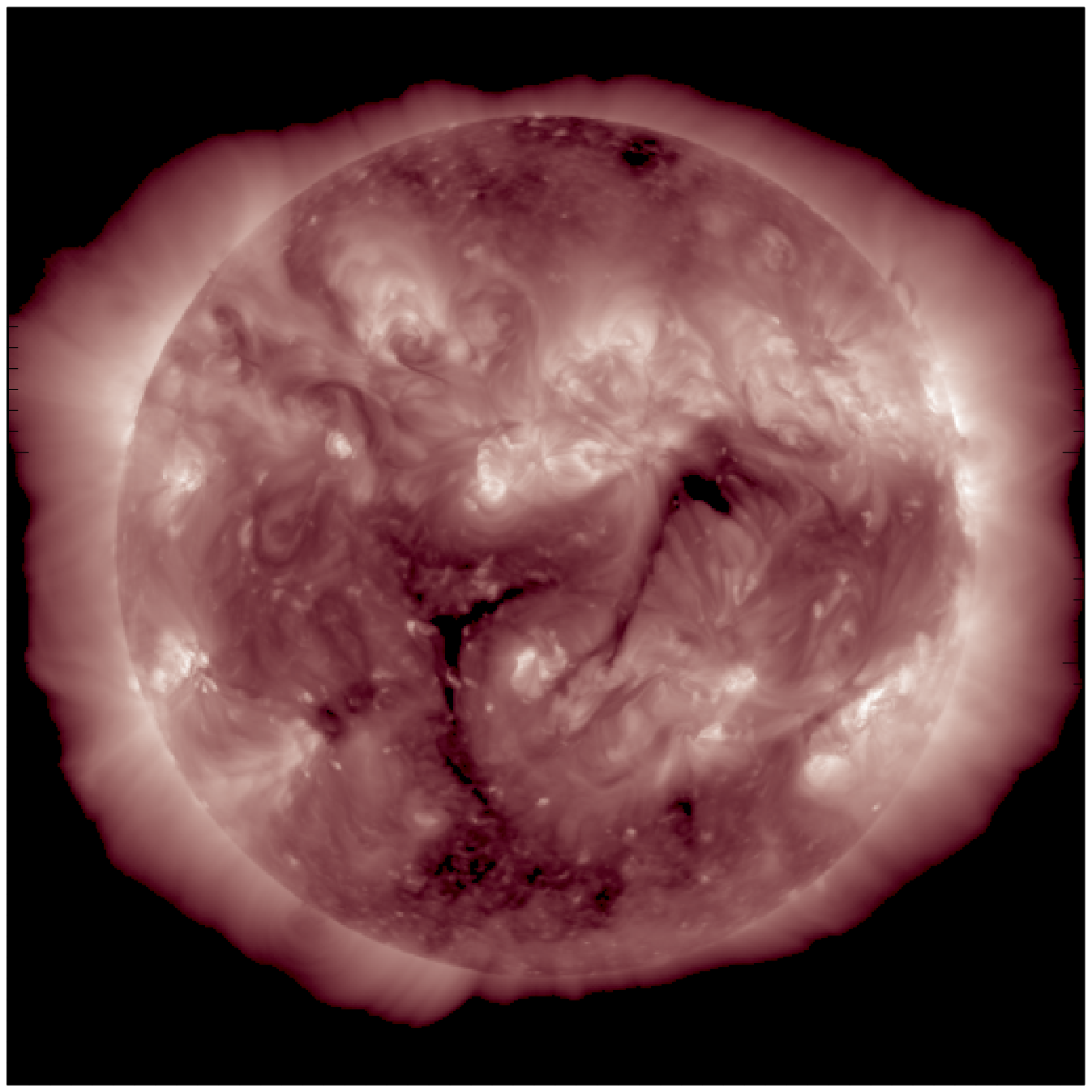}
\end{minipage}
\begin{minipage}{0.49\textwidth}
\center
\includegraphics[trim=1.5cm 1.5cm 0.5cm 2.25cm, clip=true, height=0.85\linewidth]{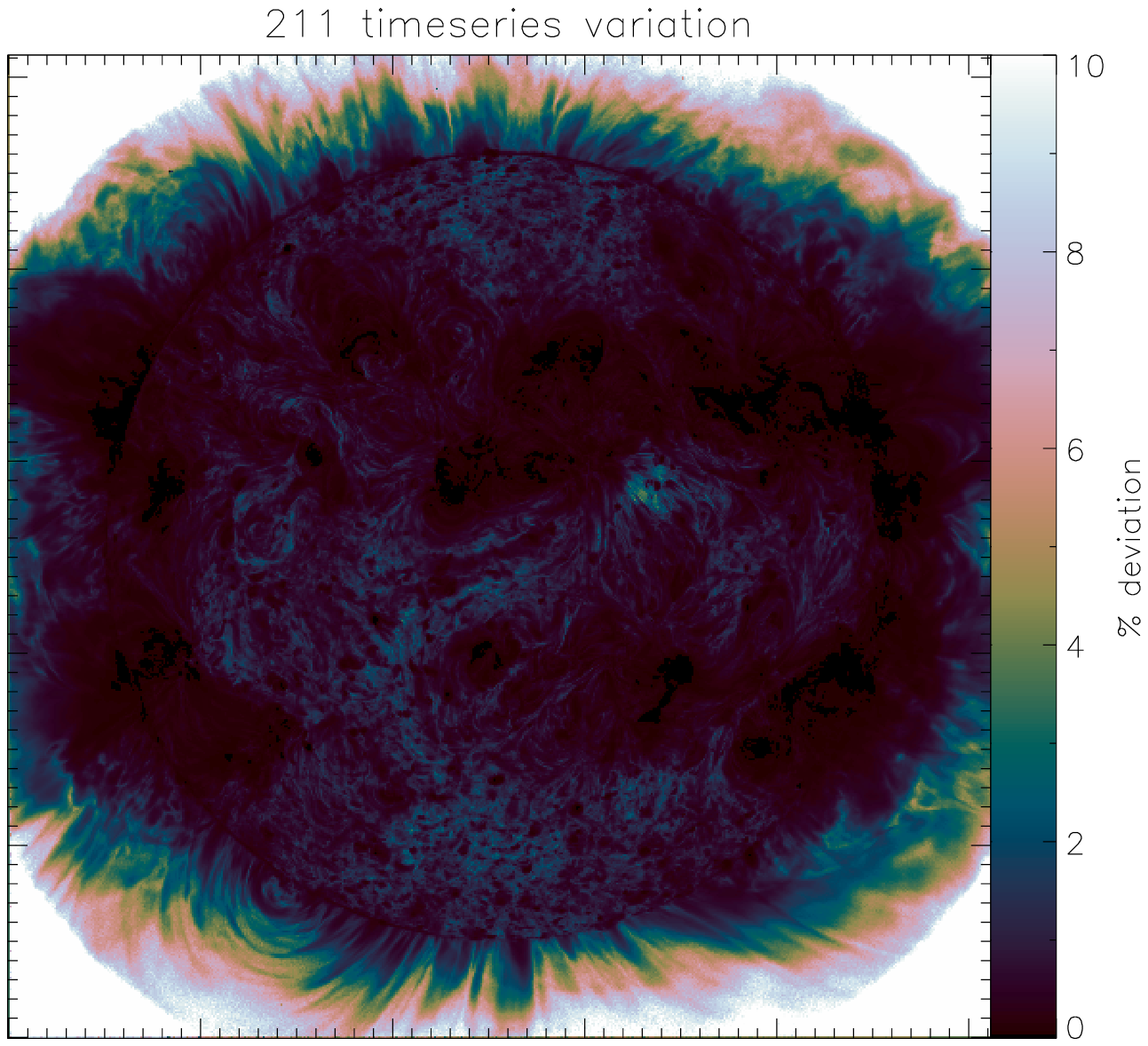}
\end{minipage}
\caption[Full disk $211$\AA\ and variability image]{Full disk solar images of \textit{left:} the eight hour integrated $211$\AA\ AIA image and \textit{right:} the standard deviation of each pixel in the $211$\AA\ image time series plotted as a percentage of the observed flux. Notice that the on-disk variation is small, especially in active regions which have high signal to noise.}
\label{fig:aia}
\end{figure*}

The full disk EUV images used for this analysis came from the Atmospheric Imaging Assembly (AIA) \citep{Lemen2012} aboard the Solar Dynamics Observatory satellite. All six coronal EUV channels (94\AA, 131\AA, 171\AA, 193\AA, 211\AA, and 335\AA) were used at one minute cadence over the course of the VLA observation. The point spread function corrections from \cite{Poduval2013a} were applied to the level 1.5 images which were then summed after rotation to a common time at the midpoint of the VLA observation to produce longer integrations and increase the signal to noise. The blurring created by small scale feature fluctuations during this long integration had no effect on our results because the bremsstrahlung prediction resulting from the EUV data was convolved with a $25\arcsec$ Gaussian beam before analysis to match the resolution of the radio observations. The EUV image sequence was also used to check for any time variability that might affect our results (Figure \ref{fig:aia}). No major time variability is present and therefore the time-integrated EUV data is appropriate for comparison with the 8--hour VLA data set.

\subsection{Helioseismic and Magnetic Imager}
Photospheric line--of--sight magnetic field measurements were obtained by the Helioseismic and Magnetic Imager (HMI) \citep{Scherrer2011} on the Solar Dynamics Observatory. These $4096\times 4096$ low--noise full sun magnetic field maps are produced every 5 minutes \citep{Schou2011a}, but only the single observation closest to 19:00 UT (the central time of the VLA observation)  was used because, like the corona, the photosphere showed very little variability during the observation window. While radio and EUV observations are sensitive to the corona and chromosphere which lie megameters above the photosphere, strong photospheric magnetic fields indicate large active regions which extend into the corona where gyroresonance emission should be the strongest. HMI magnetograms are used to qualitatively connect radio polarization measurements to the coronal magnetic field.

\section{The Differential Emission Measure}
\label{sec:DEM}
The emission measure (EM) of hydrogen in the corona is defined as:

\begin{equation}
\text{EM}=\int\limits_{0}^{l}n_{e}(s)n_{\text{H}}(s)\ \text{d}s
\label{eqn:EM_path}
\end{equation}
where $n_{e}$ is the electron number density, $n_{\text{H}}$ is the hydrogen number density, and $l$ is the path length through the optically thin medium \citep{Greenstein1953}. Since the corona is not isothermal, it is conventional to use the differential emission measure (DEM) which represents the column--integrated plasma density as a function of temperature:

\begin{equation}
\text{EM}=\int \text{DEM}(T)\ \text{d}T
\label{eqn:EM_T}
\end{equation}
with 

\begin{equation}
\text{DEM}(T)=\int \frac{\text{d}n_{e}(s)n_{\text{H}}(s)}{\text{d}T(s)}\ ds
\label{eqn:DEM}
\end{equation}
The observed intensity $I_{\lambda}$ at wavelength $\lambda$ is then given by:

\begin{equation}
I_{\lambda} = \int R_{\lambda}(T)\text{DEM}(T)\ \text{d}T
\label{eqn:intensity}
\end{equation}
where $R_{\lambda}(T)$ is the temperature response function of the instrument. This response function is dependent on both the technical details of the instrument (wavelength resolution, filter passband shape, detector response, etc.) and the atomic physics of the emitting plasma (composition, transition probabilities, occupation states, temperature sensitivity, etc.). By observing multiple emission lines with different temperature sensitivity (either as spectral lines or through narrow band imaging) it is possible to invert the system of equations (equation \ref{eqn:intensity}) to determine the DEM of the source plasma. However, with the addition of measurement errors such as Poisson noise and any instrumental effects, this system becomes under-constrained, and a precise, analytic inversion is impossible. Additionally, even if a self-consistent solution can be found, its relation to the actual emitting plasma is dependent on the atomic parameters in $R_{\lambda}(T)$ which, in the EUV, may have errors of up to $\sim 50\%$ \citep{Zanna2011}. The AIA response functions calculated using the CHIANTI package \citep{Dere2009} were used assuming coronal abundances (\S \ref{sec:Fe}) and the CHIANTI default ionization balance \citep{Boerner2012}.

\subsection{Solution Technique}
This work uses the inversion method presented in \cite{Plowman2013b}. The code is based on the original procedure presented in \cite{Hannah2012} and uses a multi-step process to invert a set of coronal flux measurements and derive the best source DEM distribution. The first pass involves a direct inversion of the input data using the instrument response functions themselves as the set of linearly--independent basis functions. This method naturally applies a minimum squared EM condition that prefers smooth solutions which tend to minimize unphysical negative EM contributions. A regularized solution with narrower basis functions is computed in order to remove the negative EM components entirely. This regularization is compared with the original solution and iteratively modified to slowly reduce the negative emission while ensuring that each step deviates by no more than the accepted $\chi ^2$ threshold. Three different regularization strengths are applied, and only if they all fail to converge are any negative EM components allowed to remain in the final solution. The primary advantage of this solution method is its speed, computing a full resolution AIA (4096$\times$4096 pixels) DEM image in about one hour on a single processor workstation \citep{Plowman2013b}. The AIA images used here are dominated by the lines of Fe VIII, IX, XII, XIV, XVIII, and XXI, which together cover the temperature range $\text{log}(T)=5.6-7.0$ corresponding to the bulk of coronal plasma.

\subsection{Full Disk DEM}
\begin{figure*}[!t]
\centering
\begin{minipage}{0.49\textwidth}
\centering
\includegraphics[trim=1cm 1cm 1cm 1cm, width=\linewidth]{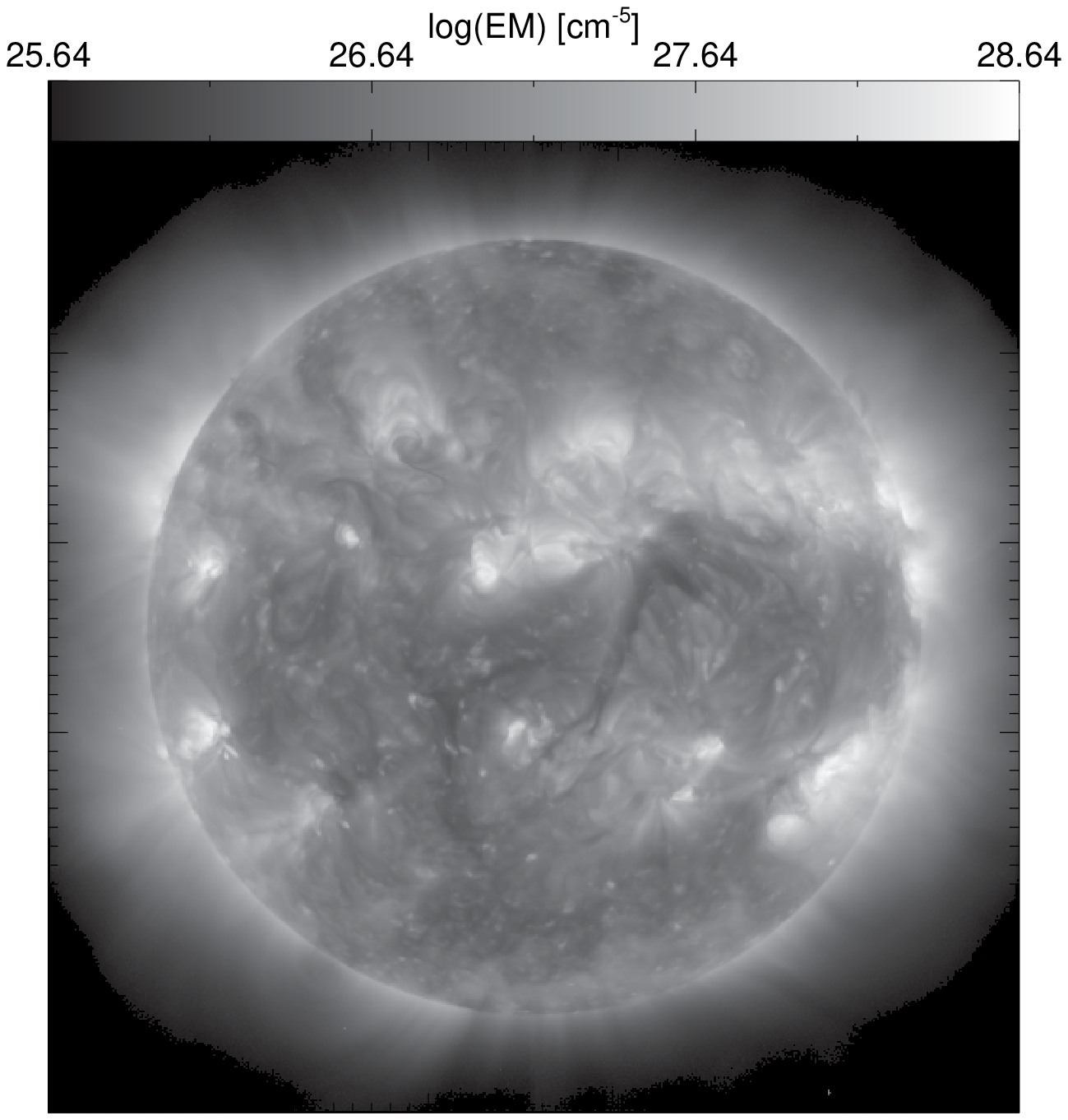}
\end{minipage}
\begin{minipage}{0.49\textwidth}
\center
\includegraphics[trim=1cm 1cm 1cm 1cm, width=\linewidth]{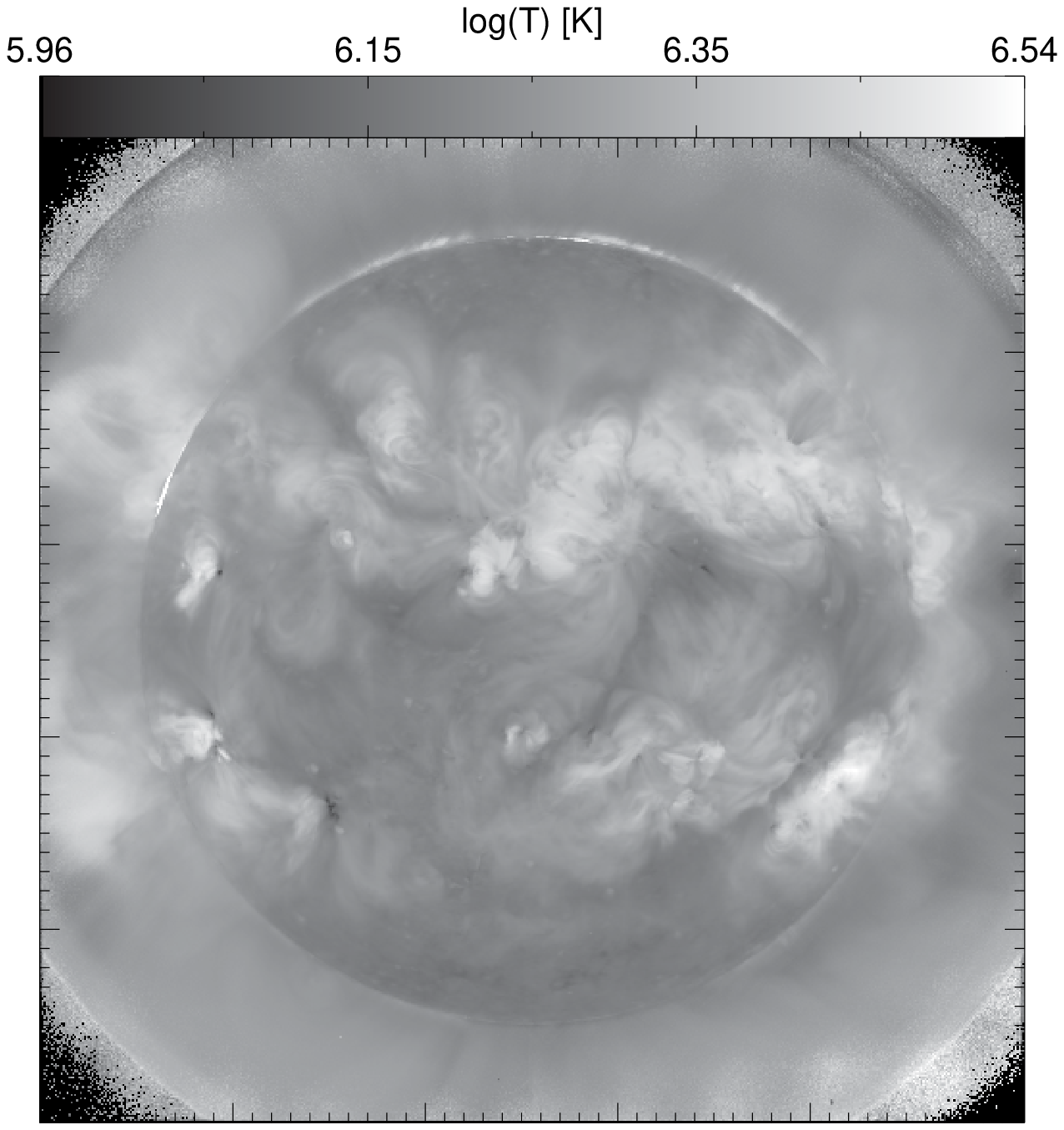}
\end{minipage}
\caption[Full disk images of the derived emission measure and median temperature]{Full disk solar images of \textit{left:} the total emission measure and \textit{right:} the emission-measure-weighted median temperature as calculated from the AIA images.}
\label{fig:dem}
\end{figure*}

Full disk representations of the calculated DEMs are shown in Figure \ref{fig:dem}. The left image shows the total emission measure which was obtained by integrating the DEM in each pixel over the temperature axis. This EM dominates the contribution to the final bremsstrahlung prediction because it is linear in the integral in equation \ref{eqn:Bremsstrahlung} and has over two orders of magnitude variation on the solar disk. On the right is an image of the emission-measure-weighted median temperature calculated from the derived full disk DEM. This map shows some small discontinuities along the east and north limb (which appear saturated in the image) but these regions play no role in the analysis as described in \S \ref{sec:analysis}. The temperature impacts the bremsstrahlung emission as $T^{-0.5}$ \citep{Dulk1985} in the integral over temperature (Equation \ref{eqn:Bremsstrahlung}), and, because the median temperature varies by only half an order of magnitude across the solar disk, the temperature variation plays a relatively minor role in the predicted radio fluxes compared to the emission measure.

\section{Analysis}
\label{sec:analysis}
\begin{figure*}[!t]
\centering
\begin{minipage}{0.49\textwidth}
\centering
\includegraphics[trim=0cm 0cm 1cm 1cm, width=\linewidth]	{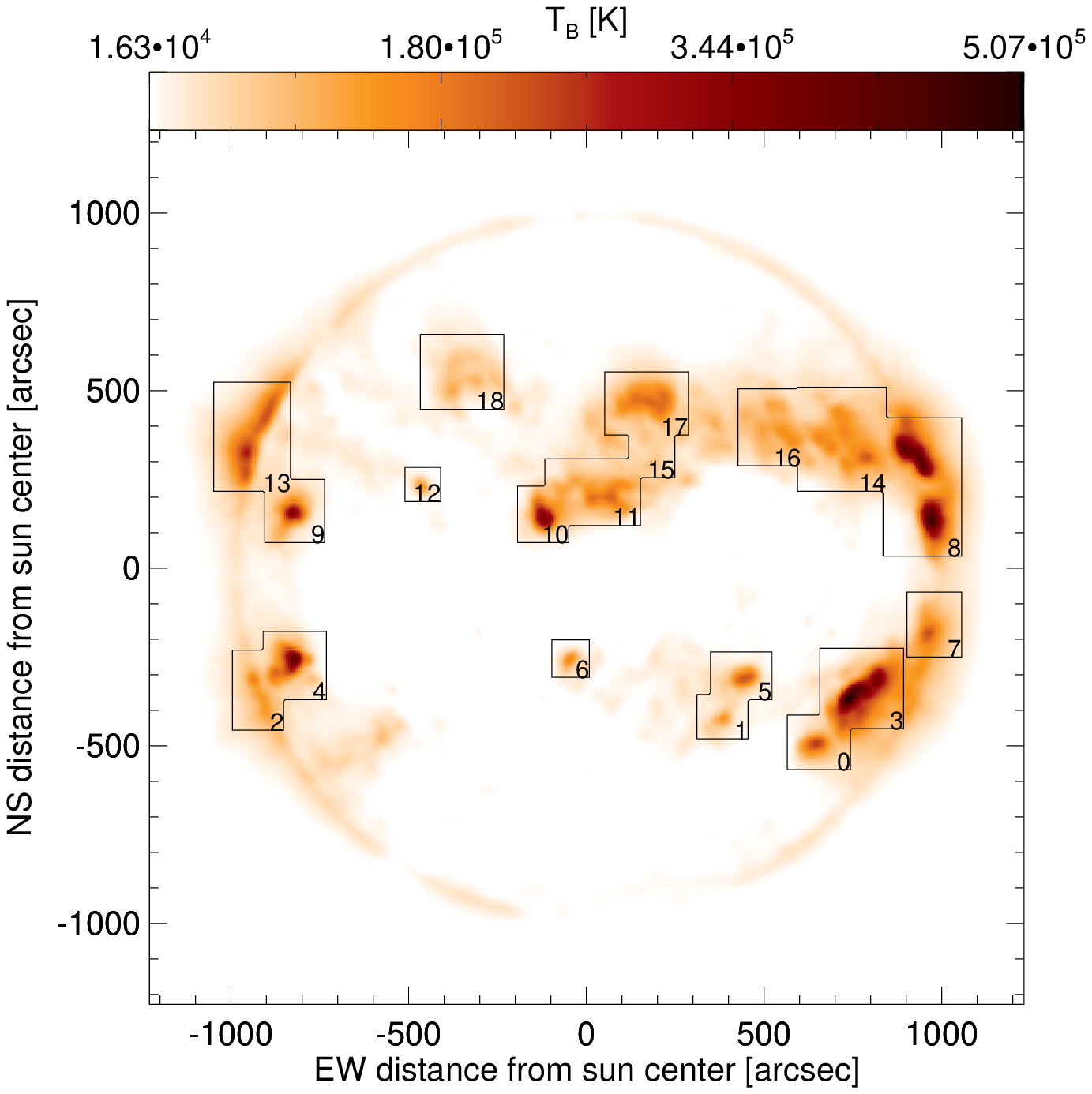}
\end{minipage}
\begin{minipage}{0.49\textwidth}
\center
\includegraphics[trim=0cm 0cm 1cm 1cm, width=\linewidth]	{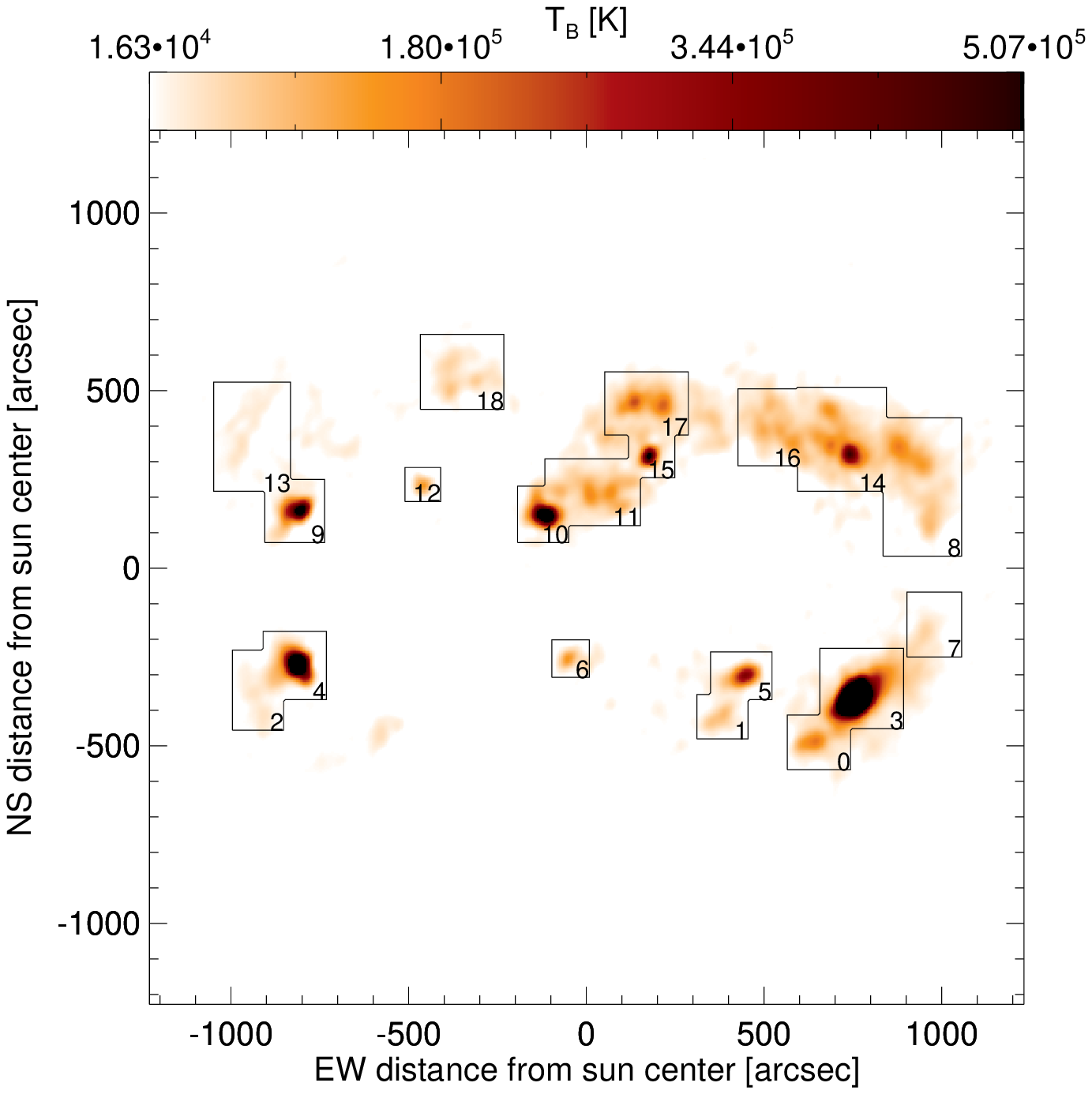}
\end{minipage}
\caption[Full disk images of the sun from bremsstrahlung prediction and radio observation.]{Full disk solar images on the same brightness temperature scale of \textit{left:} the optically thin bremsstrahlung prediction and \textit{right:} the $2.783$ GHz observation. Individually analyzed regions are boxed and numbered.}
\label{fig:full_disk}
\end{figure*}

An image of the expected coronal component of the bremsstrahlung emission was calculated using the DEMs computed from the AIA images and equation \ref{eqn:Bremsstrahlung}, assuming a coronal abundance for Fe discussed in \S \ref{sec:Fe}. This image was computed at the full AIA resolution and then convolved with a $25\arcsec$ full width half maximum Gaussian and down--binned to match the resolution of the radio observations. The brightness temperature of the bremsstrahlung prediction and radio observation are shown in Figure \ref{fig:full_disk} on the same scale, with individual regions outlined and labelled. Notice that while the overall morphologies agree quite well, the observation has large, high temperature emission peaks from the centres of many of the regions which are absent from the bremsstrahlung prediction. Additionally, the bremsstrahlung prediction has significantly more faint emission surrounding the active regions than was observed. It is important to emphasize that the prediction image is based on the AIA observations and therefore it will vary from the observed optically thin bremsstrahlung wherever the EUV is sensitive to different plasma than the radio.

The average F$_{10.7}$ flux measured by the official Solar Monitoring Program in Penticton, Canada, at $18$, $20$, and $22$ UT on 2011 December 9 was $143.5\pm 1.2$ sfu. Subtracting the quiet sun background of $67.2\pm 2.1$ sfu (the $65.2\pm 2.0$ sfu constant solar minimum level scaled to a Sun-Earth separation of $0.985$ AU) leaves an observed variable F$_{10.7}$ component of $76.3\pm 2.4$ sfu. The total F$_{10.7}$ flux from the predicted bremsstrahlung image was $77.7\pm 0.1$ sfu, which compares well with the observed variable flux. This suggests that gyroresonance emission is not distorting the F$_{10.7}$ flux significantly since the optically thin bremsstrahlung component can account for all of the variable F$_{10.7}$ on this day. However, this conclusion ignores several complicating details which are discussed in \S \ref{sec:discussion}.

\subsection{Region comparison}
\begin{figure*}[!t]
\centering
\begin{subfigure}{0.375\paperwidth}
	\includegraphics[trim=0.25cm 0.25cm 1cm 0.25cm, clip=true, width=\linewidth]		{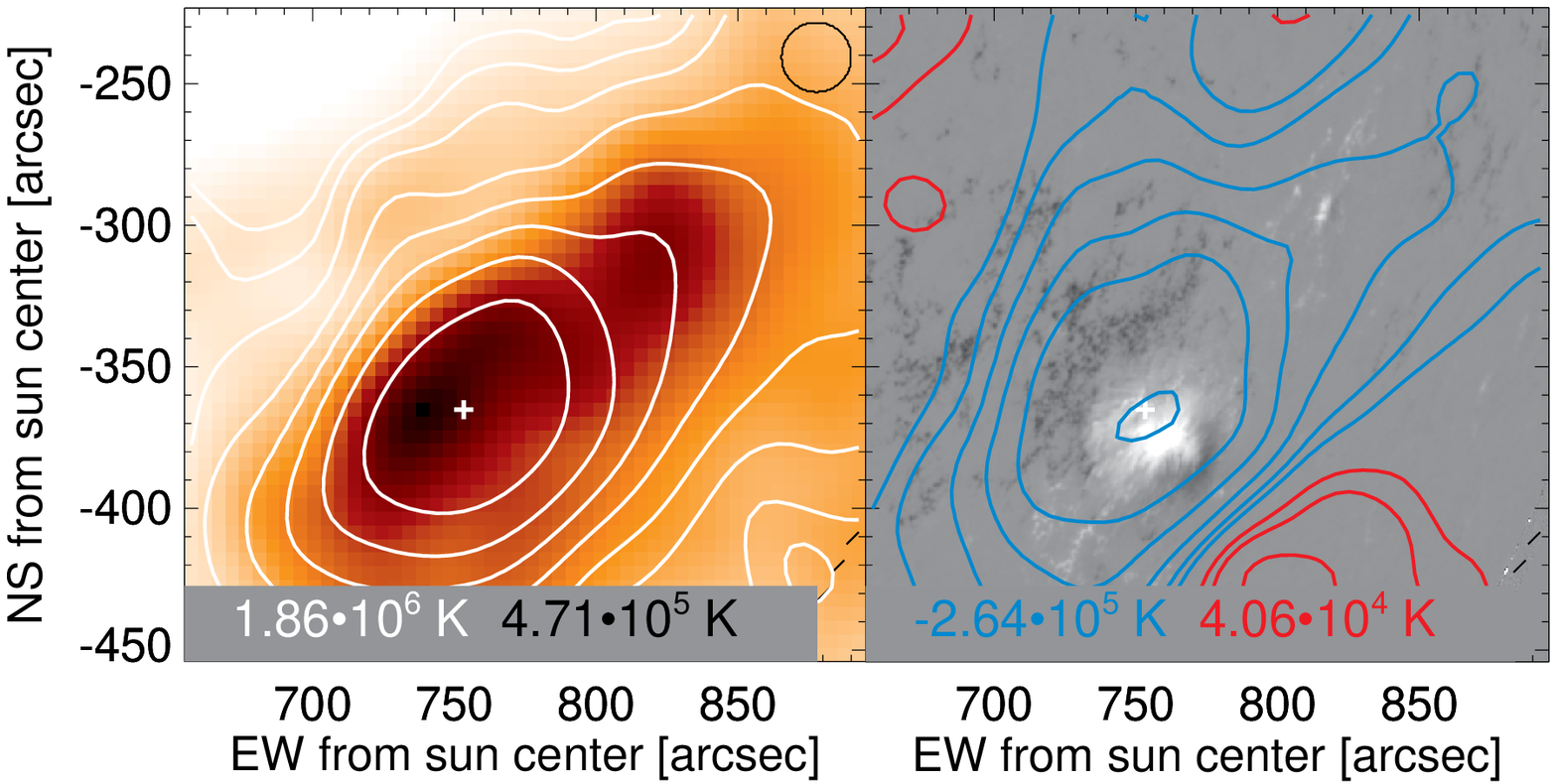}
	\caption{Region 3}
	\label{fig:region3}
\end{subfigure}
\begin{subfigure}{0.375\paperwidth}
	\includegraphics[trim=0.25cm 0.25cm 1cm 0.25cm, clip=true, width=\linewidth]{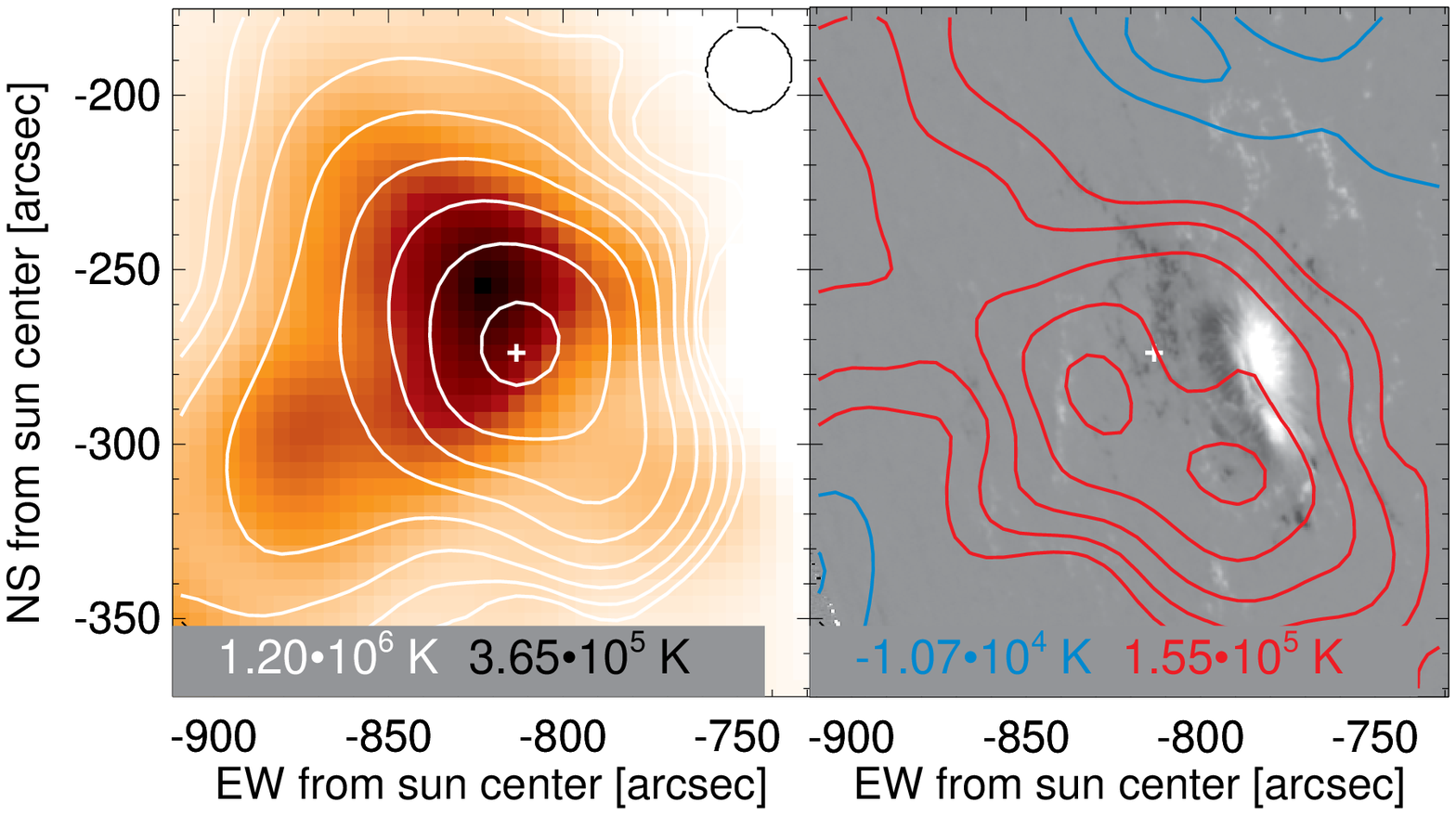}
	\caption{Region 4}
	\label{fig:region4}
\end{subfigure}

\vspace{0.25cm}

\begin{subfigure}{0.375\paperwidth}
	\includegraphics[trim=0.25cm 0.25cm 1cm 0.25cm, clip=true, width=\linewidth]{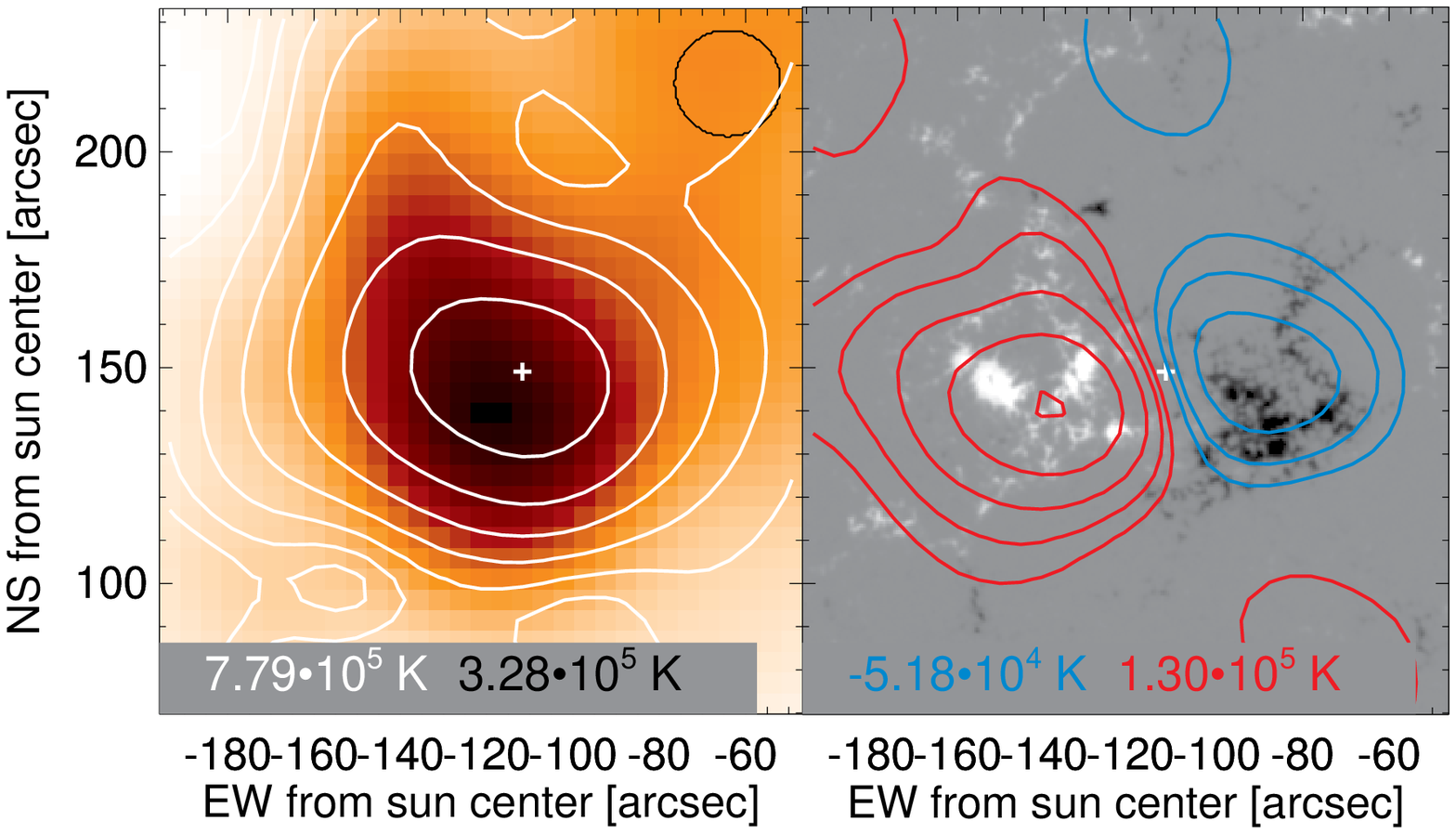}
	\caption{Region 10}
	\label{fig:region10}
\end{subfigure}
\begin{subfigure}{0.375\paperwidth}
	\includegraphics[trim=0.25cm 0.25cm 1cm 0.25cm, clip=true, width=\linewidth]{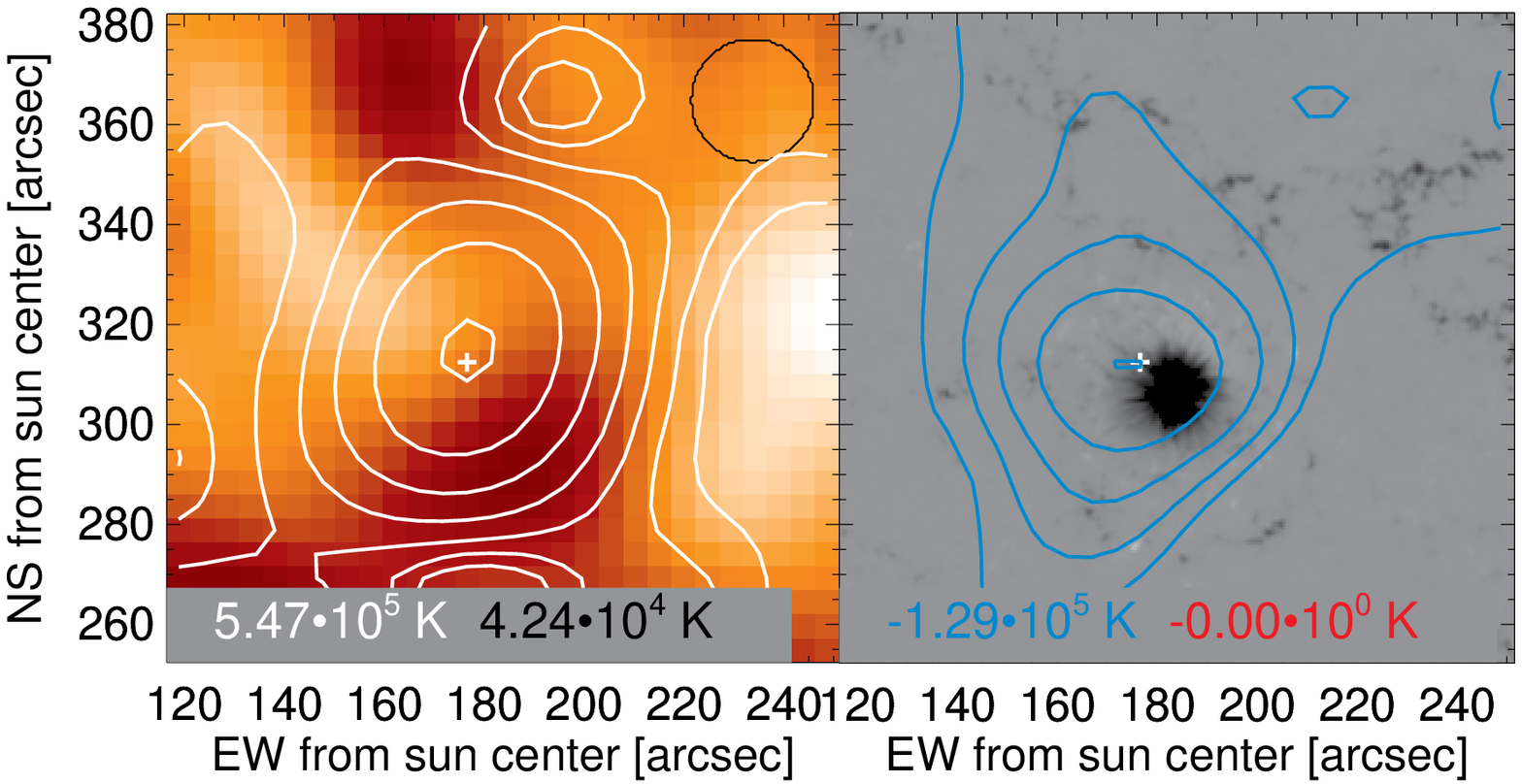}
	\caption{Region 15}
	\label{fig:region15}
\end{subfigure}

\caption[Cut outs of regions of interest]{Individual regions of interest which each have a peak polarization brightness temperature greater than $10^{5}$ K. The left pane shows the bremsstrahlung prediction as an inverted heat map with the radio intensity over-plotted with white contours at $(8,16,32,64,128,256,512,1024)\times10^{3}$ K. The right pane shows the photospheric line-of-sight magnetic field in gray scale with the radio polarization brightness plotted with contours (blue for left hand polarization and red for right hand polarization) at $\pm(8,16,32,64,128,256,512)\times10^{3}$ K. The small black circle in the top right corner of the left pane is the approximate beam size and the white plus signs mark the location of peak observed radio intensity. The left pane also lists the peak brightness temperature of the radio observation (white) and the bremsstrahlung prediction (black) while the right pane lists the minimum (left hand) and maximum (right hand) polarization brightness temperatures.}
\label{fig:cutouts}
\end{figure*}

The full disk images in Figure \ref{fig:full_disk} cannot be compared quantitatively because of the failure to restore the flat background disk (which is in any case absent from the EUV images) to the F$_{10.7}$ radio image. Instead, individual regions on the Sun for which the imaging is reliable were analyzed. The bremsstrahlung prediction, radio intensity, circular polarization, and photospheric line--of--sight magnetic field strength in each region were compared in order to determine if gyroresonance emission was present. These comparisons for the regions with the largest observed polarizations are shown in Figure \ref{fig:cutouts}. In region 3, the general morphologies of the observed and predicted active regions agree very well despite the much higher observed radio brightness temperature than predicted by the EUV data. Additionally, the circular polarization signal is large above the strong photospheric magnetic fields, as expected. The remaining three regions shown in the figure each display varying levels of morphological deviation between the bremsstrahlung prediction and radio observation, suggesting that there are gyroresonance sources offset from the peak bremsstrahlung emission (because optical depth effects in the bremsstrahlung sources will not produce offsets). In region 4, the peak observed emission is shifted to the southwest by about $20\arcsec$, aligning with the observed polarization signal. The observation of region 10 extends east and west farther than the predicted emission, aligning with the polarization signal directly above the east-west oriented photospheric magnetic fields. Region 15 shows a strongly polarized radio source above an isolated sunspot with no corresponding predicted bremsstrahlung source. These regions show that the VLA resolution is sufficient to extract active region details in both the intensity and the individual polarization channels, allowing for the identification of large gyroresonance sources simply from the polarization and morphological inconsistencies.

\begin{figure*}[!t]
\centering
\begin{minipage}{0.49\textwidth}
\centering
\includegraphics[trim=0.5cm 0.25cm 0.5cm 0.5cm, clip=true, width=\linewidth]{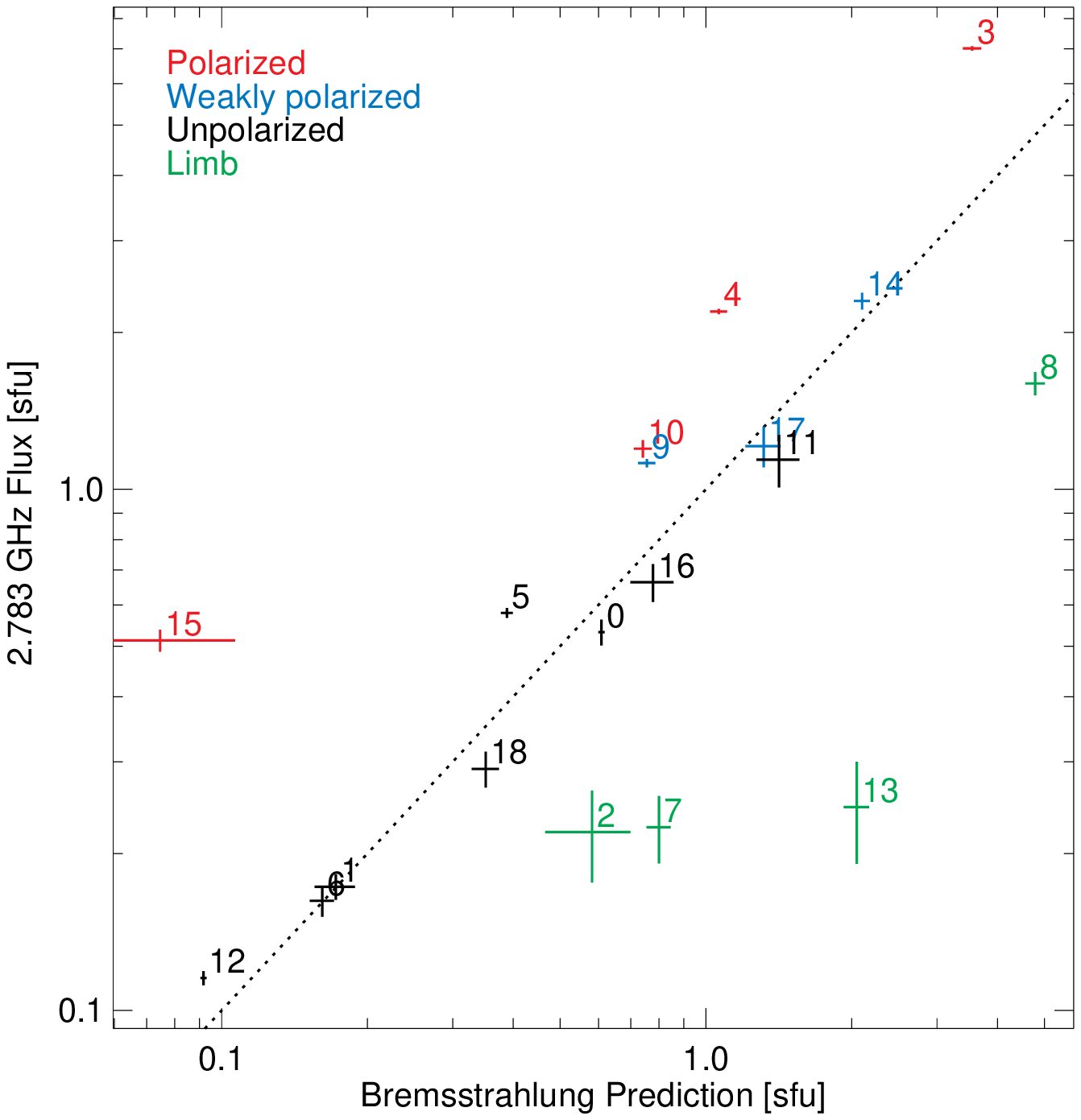}
\end{minipage}
\begin{minipage}{0.49\textwidth}
\center
\includegraphics[trim=0.5cm 0.25cm 0.5cm 0.5cm, clip=true, width=\linewidth]{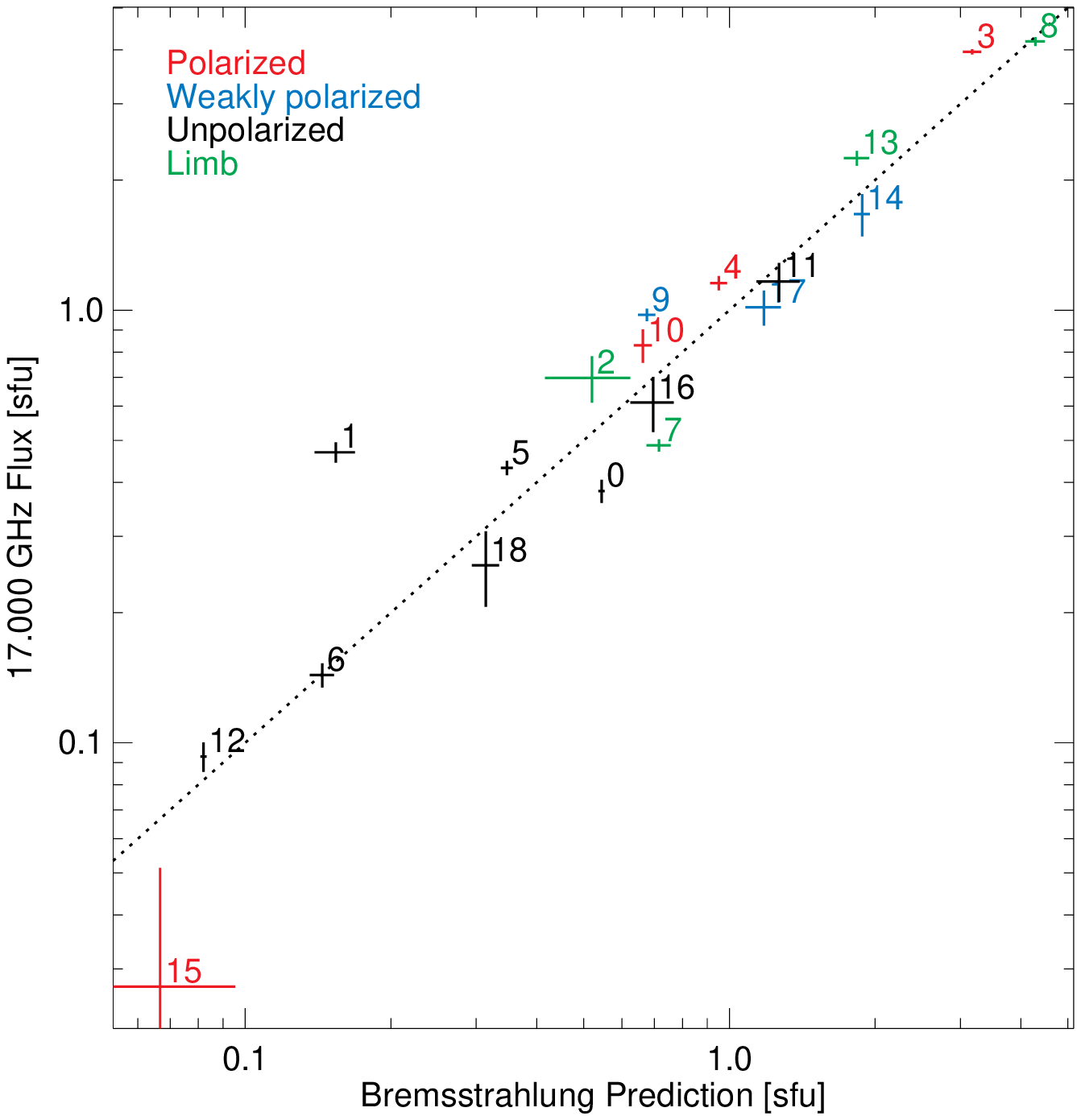}
\end{minipage}
\caption[Flux comparison in individual regions]{Total flux observed in each region at \textit{left:} $2.783$ GHz and \textit{right:} $17$ GHz is plotted against the total predicted optically thin bremsstrahlung emission. The dotted lines indicate where the bremsstrahlung prediction equals the observed flux. Regions which lie above the line have more observed flux than is predicted, suggesting gyoresonance emission. Regions which lie below the line are non-physical and indicate more predicted optically thin bremsstrahlung emission than the total observed flux. Regions labelled in red have peak polarization brightness temperatures of $T_{B}\geq 10^{5}$K, blue regions have $5\times 10^{4}\leq T_{B}\leq 10^{5}$K, black regions have $T_{B}\leq 5\times 10^{4}$K, and green regions lie above the solar limb.}
\label{fig:flux_absolute}
\end{figure*}

In order to isolate the active region fluxes from any larger-scale background, identical background--subtraction approaches were used in both the bremsstrahlung prediction and the radio observation to allow quantitative comparison between the data sets. This involved using the solar disk around the border of each region to estimate the disk emission within the region itself. This was done for concentric borders up to $3\arcmin$ outside each region, using the average result to calculate the region flux and the variation in the total region flux as a measure of the uncertainty. Note that while these uncertainties are quoted for the remainder of the paper, the systematic uncertainties associated with the DEM calculation and the uncertainty in the VLA flux calibration may be much larger (as discussed in \S \ref{sec:uncertainties}). The flux from each region is plotted in Figure \ref{fig:flux_absolute} where the regions have been classified based on their maximum polarization brightness temperature and their proximity to the solar limb. 

\textbf{We emphasize peak circular polarization brightness temperature rather than degree of polarization because the effective noise level in the latter is very high. This is due to both the noise level in Stokes I and V and our inability to fully restore the disk emission in total intensity which greatly affects the low intensity regions. In addition, the large beam size at 2.8 GHz results in smearing between any smaller gyroresonance sources and the more extended bremsstrahlung emission. In some cases (notably region 10), oppositely polarized gyroresonance sources overlap within the $25\arcsec$ beam size and cancel, resulting in artificially low polarization. All the regions with strong circular polarization show degrees of polarization greater than 30\%, but this is likely an underestimate.}

\section{Discussion}
\label{sec:discussion}
\subsection{Full disk}
A dramatic feature of Figure \ref{fig:full_disk} is that the EUV--predicted radio fluxes of the regions at the solar limb are all well in excess of their counterparts in the radio image. This effect can also be seen in the left panel of Figure \ref{fig:flux_absolute} where the limb regions are plotted as green points which all lie in the non-physical regime. The total predicted bremsstrahlung flux from the limb regions was $8.2\pm 0.3$ sfu whereas only $2.3\pm 0.1$ sfu was observed. Correcting for this $5.9\pm 0.3$ sfu difference in the limb flux suggests that there should be $71.8\pm 0.3$ sfu of optically thin coronal bremsstrahlung in the F$_{10.7}$ signal. Comparison to the observed $76.3\pm 2.4$ sfu variable component now suggests the presence of a small amount of gyroresonance emission. Note that this is a conservative correction for the discrepancy at the limb because it does not account for any of the limb emission outside the main active regions.

The chromosphere provides the optically thick background for both the radio and EUV observations and therefore sets the height of the visible solar limb, however this height is frequency dependent. The effective solar limb at $2.8$ GHz is around $30\arcsec$ above the solar photosphere \citep{Gary1996}, while the height of the effective solar limb at EUV wavelengths is only a few arcseconds above the photosphere \citep{Auchere1998,Zhang1998}. It is believed that the extra height of the radio limb is due to cool filamentary chromospheric material (such as spicules) that extends into the solar corona and can provide extra opacity at the limb. We interpret the depressed radio signal from the limb regions as the occultation of emission originating behind the chromospheric limb. This has interesting implications for F$_{10.7}$ as an EUV proxy because the difference in limb altitude may cause the obscuration of a significant fraction of the solar emission in the radio which is visible in the EUV. In this case, plasma that produced at least $7.7\pm 0.5\%$ of the variable F$_{10.7}$ component and which was visible in EUV was not observed in the radio. Not only does this complicate comparison of F$_{10.7}$ and EUV fluxes, it also causes an offset in time-series comparisons because EUV sources will become visible before corresponding F$_{10.7}$ sources rotate into view, and will remain visible after F$_{10.7}$ sources rotate behind the limb.

\subsection{Individual regions}
Bremsstrahlung emission is usually weakly polarized and gyroresonance sources are often strongly polarized, therefore those regions which had highly polarized emission were suspected gyroresonance sources. In Figure \ref{fig:flux_absolute} the observed regions were categorized based on the peak polarization brightness temperature. Regions with a peak polarization brightness temperature $T_{B}\geq 10^{5}$ K at $2.783$ GHz were deemed to have strong polarization and all lie well above the bremsstrahlung line, therefore confirming that they have significant gyroresonance contribution. Regions with $5\times 10^{4}\leq T_{B}\leq 10^{5}$ K were classified as weakly polarized, possibly containing gyroresonance sources, while regions with $T_{B}\leq 5\times 10^{4}$ K were deemed to be insufficiently polarized and were unlikely to contain significant gyroresonance sources. These weakly polarized and unpolarized regions all have fluxes very close to the bremsstrahlung predictions, with only two (regions 5 and 9) showing significant observed radio excess. The total radio excess which is interpreted as the gyroresonance contribution (although this could also be due in part to the bremsstrahlung becoming optically thick) was $5.6\pm 0.2$ sfu in the regions with strong polarization, $0.3\pm 0.2$ sfu in the weakly polarized regions, and $0.3\pm 0.2$ sfu in the unpolarized regions. More than $60\%$ of the total gyroresonance emission originates in region 3, the largest disk active region.

This analysis suggests that $6.2\pm 0.3$ sfu or $8.1\pm 0.5\%$ of the variable F$_{10.7}$ signal recorded on 2011 December 9 resulted from gyroresonance emission. While this is a small percentage it is well above the precision of the Penticton F$_{10.7}$ measurements and could be sufficient to account for the known $\sim10\%$ density errors resulting from F$_{10.7}$ driven thermosphere models \citep{Bowman2008}. It is important to note that this gyroresonance emission, if it were constant (or even a constant fraction of the variable component), would have little impact on the use of F$_{10.7}$ as an EUV proxy. However, since the gyroresonance contribution is dominated by the largest active region, we speculate that it is likely to be a much larger fraction of F$_{10.7}$ at times of high solar activity and this could affect the use of F$_{10.7}$ as an EUV proxy.

\subsection{Coronal Iron Abundance}
\label{sec:Fe}
A straightforward result of this analysis is confirmation that the iron abundance in the corona is $N_{\text{Fe}}/N_{\text{H}}= 1.26\times 10^{-4}$, an enhancement of about a factor of $4$ over the photospheric value. Two independent results which depend on the coronal value of $N_{\text{Fe}}/N_{\text{H}}$ confirm this conclusion: the total bremsstrahlung flux derived from the AIA data matches the coronal contribution to F$_{10.7}$ derived by subtracting the solar minimum flux from the measured F$_{10.7}$ flux during the observation; and the fluxes of the $17$ GHz active regions, which are known to be well calibrated by matching the NoRH full-disk flux to the calibrated NoRP patrol measurement, are consistent with the predictions from AIA data. The inferred DEM of H, to which the predicted bremsstrahlung radio flux is proportional, depends inversely on the assumed iron abundance: if $N_{\text{Fe}}/N_{\text{H}}$ were to be photospheric, the F$_{10.7}$ prediction based on the AIA data for this day would be $67.2+(4*77.7)=378$ sfu, rather than the measured $143.5\pm 1.2$ sfu. These results are consistent with \cite{White2000a} who found an iron abundance of $N_{\text{Fe}}/N_{\text{H}}=1.56\times10^{-4}$ with approximate $20\%$ errors by comparing radio observations with bremsstrahlung predictions from EUV DEMs in a single active region. Note that our assumption that the solar minimum flux has no significant coronal contribution disagrees with the model of \cite{Zirin1991}, but \cite{Landi2003, Landi2008} carried out a careful comparison of the solar minimum radio spectrum with a DEM derived from UV and EUV data and in their results the F$_{10.7}$ solar minimum flux has only a small coronal contribution.

Features other than coronal emission from active regions, including flares (which may be dominated by evaporated chromospheric material) and energetic particles, have shown different abundances, ranging from 1.2 \citep{Meyer1985} to 13.1 \citep{Reames1999} times the photospheric level. Our results suggests that an iron enhancement of $4$ is generally appropriate for coronal active regions: we will pursue this result further in a future study.

\subsection{Uncertainties}
\label{sec:uncertainties}
It should again be noted that the systematic uncertainties involved in this analysis are much greater than the statistical errors quoted above. Systematics such as the iron abundance and improper calibration of the VLA $20$ dB attenuators cause constant offsets and therefore affect the overall agreement of the disk fluxes (although this does not apply to the well-calibrated $17$ GHz data). The effects of these constant offsets is minimized by normalizing each region to the local disk background as described in section \ref{sec:analysis}. However, based on the errors in coronal abundance studies as well as differences between the flux observed with the VLA and the official record measured at Penticton, we expect both of these error sources to be on the order of $20\%$.

Uncertainties in the DEM solutions could have spatially variable errors depending on the underlying plasma parameters. We expect these errors to be on the order of $10\%$ or less but it is difficult to quantify the extent to which deviations between the derived DEM and the ground truth plasma parameters change the results due to the non-linear influence of the DEM on the bremsstrahlung prediction. However, the comparison of the bremsstrahlung prediction to the $17$ GHz Nobeyama observation in Figure \ref{fig:flux_absolute} shows no clear correlation of deviation with active region size. This suggests that there is no systematic bias and that any pixel scale statistical errors in the DEM are washed out when integrating over an entire active region.

No attempt was made to account for the underestimation of gyroresonance emission because of the predicted bremsstrahlung emission originating from below the optically thick gyroresonance layer. If the most extreme case is assumed, that all of the observed radio emission from the strongly polarized regions resulted from gyroresonance emission (i.e., all of the predicted bremsstrahlung emission occurred below the optically thick gyroresonance layer), then the total gyroresonance flux from these regions would be $10.9\pm 0.1$ sfu. This is a generous upper limit which, while it does allow a possible factor of two difference in the gyroresonance flux, still restricts the total gyroresonance to less than $15\%$ of the variable F$_{10.7}$ component.

\section{Conclusion}
\label{sec:conclusion}
Understanding the sources of the solar F$_{10.7}$ flux is important if it is to be used reliably as an EUV proxy in thermosphere/ionosphere models. By comparing a full disk VLA observation with the F$_{10.7}$ bremsstrahlung emission predicted from DEMs calculated with AIA images, we find that $8.1\pm 0.5\%$ of the variable F$_{10.7}$ flux on 2011 December 9 can be attributed to the gyroresonance mechanism. This gyroresonance contribution does not directly correlate with solar EUV flux and therefore should be removed from F$_{10.7}$ when it is used as an EUV proxy. While this is a small fraction of the F$_{10.7}$ signal, it is commensurate with the density errors in current ionospheric modelling efforts.

We also identify unexpected occultation due to the optically thick chromosphere of F$_{10.7}$ flux originating from behind the solar limb. It appears that this effect could cause systematic errors in F$_{10.7}$ time series at the active region scale (on the order of $10$ sfu) at one day time scales. To our knowledge, this effect has not previously been considered as a possible complication when comparing F$_{10.7}$ to EUV emission. For our observation, the on-disk gyroresonance excess ($6.2\pm 0.3$ sfu) almost perfectly cancels the off disk paucity (minimum of $5.9\pm 0.3$ sfu). There is no reason for these two effects to be correlated except in the general sense that both are likely to vary with the general level of activity, and consequently it seems unlikely that they will generally offset each other as well as they do here.

The details of these results depend strongly on the coronal iron abundance which is inherent in the calculation of the DEM. By comparing the bremsstrahlung prediction with the coronal contribution to the F$_{10.7}$ measurement and with the Nobeyama $17$ GHz images, we confirm a coronal iron abundance of $N_{\text{Fe}}/N_{\text{H}}= 1.26\times 10^{-4}$ \citep[the standard coronal abundance in the CHIANTI database; ][]{Dere1997, Landi2013} which is used in the calculation of AIA temperature response functions. These results are subject to various potential systematic error sources which are difficult to quantify, but have estimated accuracies on the order of $20\%$.

No general statements about the effects of gyroresonance and limb corrections on the F$_{10.7}$ index can be made at this time because a single measurement is insufficient to characterize their temporal variability. Additionally, based on the untested systematic biases from the VLA calibration complications and the confusion regarding the altitude of optically thick gyroresonance layers, caution should be taken when considering these results. The temporal variability of the gyroresonance fraction and the effects of gyroresonance emission altitude will be addressed further in future studies for which the data have already been collected.

\acknowledgements
\textit{Acknowledgements:} S.W. would like to acknowledge valuable conversations with Ken Tapping. The National Radio Astronomy Observatory is a facility of the National Science Foundation operated under cooperative agreement by Associated Universities, Inc. We thank the Nobeyama Solar Radio Observatory for providing ready access to their data. Data supplied courtesy of the SDO/HMI and SDO/AIA consortia. SDO is the first mission to be launched for NASA's Living With a Star (LWS) Program. CHIANTI is a collaborative project involving George Mason University, the University of Michigan (USA) and the University of Cambridge (UK). This research has been made possible with funding from AFOSR FA9550-15-1-0014, NSF Career Award \#1255024, New Mexico Space Grant (a NASA EPSCoR branch), and PAARE NSF:0849986.

\bibliography{library,books,manually_added}

\begin{thebibliography}{}
\expandafter\ifx\csname natexlab\endcsname\relax\def\natexlab#1{#1}\fi

\bibitem[{Asplund {et~al.}(2009)Asplund, Grevesse, Sauval, \&
  Scott}]{Asplund2009}
Asplund, M., Grevesse, N., Sauval, a.~J., \& Scott, P. 2009, Annual Review of
  Astronomy and Astrophysics, 47, 481

\bibitem[{Auchere {et~al.}(1998)Auchere, Boulade, Koutchmy, Smartt,
  Delaboudiniere, Georgakilas, Gurman, \& Artzner}]{Auchere1998}
Auchere, F., Boulade, S., Koutchmy, S., {et~al.} 1998, Astronomy \&
  Astrophysics, 336L, 57

\bibitem[{Bastian \& Dulk(1988)}]{Bastian1988a}
Bastian, T.~S., \& Dulk, G.~A. 1988, in Solar and Stellar Coronal Structure and
  Dynamics, ed. R.~C. Altrock, 386--391

\bibitem[{Bhatnagar \& Mitra(1966)}]{Bhatnagar1966}
Bhatnagar, V.~P., \& Mitra, A.~P. 1966, Journal of the Atmospheric Sciences,
  23, 233

\bibitem[{Boerner {et~al.}(2012)Boerner, Edwards, Lemen, Rausch, Schrijver,
  Shine, Shing, Stern, Tarbell, Title, Wolfson, Soufli, Spiller, Gullikson,
  McKenzie, Windt, Golub, Podgorski, Testa, \& Weber}]{Boerner2012}
Boerner, P., Edwards, C., Lemen, J., {et~al.} 2012, Sol. Phys., 275, 41

\bibitem[{Bouwer(1992)}]{Bouwer1992}
Bouwer, S.~D. 1992, Solar Physics, 142, 365

\bibitem[{Bowman {et~al.}(2008)Bowman, {Kent Tobiska}, Marcos, \&
  Valladares}]{Bowman2008}
Bowman, B.~R., {Kent Tobiska}, W., Marcos, F.~a., \& Valladares, C. 2008,
  Journal of Atmospheric and Solar-Terrestrial Physics, 70, 774

\bibitem[{Chen {et~al.}(2011)Chen, Liu, \& Wan}]{Chen2011a}
Chen, Y., Liu, L., \& Wan, W. 2011, Journal of Geophysical Research, 116,
  A04304

\bibitem[{Covington(1947)}]{Covington1947}
Covington, A.~E. 1947, Nature, 159, 405

\bibitem[{Covington(1948)}]{Covington1948}
---. 1948, Proceedings of the IRE, 36, 454

\bibitem[{Covington(1951)}]{Covington1951}
---. 1951, Journal of the Royal Astronomical Society of Canada, 45, 15

\bibitem[{Covington(1969)}]{Covington1969}
---. 1969, Journal of the Royal Astronomical Society of Canada, 63, 125

\bibitem[{Covington {et~al.}(1955)Covington, Medd, Harvey, \&
  Broten}]{Covington1955}
Covington, A.~E., Medd, W.~J., Harvey, G.~A., \& Broten, N.~W. 1955, Journal of
  the Royal Astronomical Society of Canada, 49, 235

\bibitem[{Craig \& Brown(1976)}]{Craig1976}
Craig, I. J.~D., \& Brown, J.~C. 1976, Astronomy \& Astrophysics, 49, 239

\bibitem[{Deng {et~al.}(2013)Deng, Li, Zheng, \& Cheng}]{Deng2013a}
Deng, L.~H., Li, B., Zheng, Y.~F., \& Cheng, X.~M. 2013, New Astronomy, 23-24,
  1

\bibitem[{Dere {et~al.}(1997)Dere, Landi, Mason, {Monsignori Fossi}, \&
  Young}]{Dere1997}
Dere, K.~P., Landi, E., Mason, H.~E., {Monsignori Fossi}, B.~C., \& Young,
  P.~R. 1997, Astronomy and Astrophysics Supplement Series, 125, 149

\bibitem[{Dere {et~al.}(2009)Dere, Landi, Young, {Del Zanna}, Landini, \&
  Mason}]{Dere2009}
Dere, K.~P., Landi, E., Young, P.~R., {et~al.} 2009, Astronomy \& Astrophysics,
  498, 915

\bibitem[{{Dudok de Wit} {et~al.}(2014){Dudok de Wit}, Bruinsma, \&
  Shibasaki}]{DudokdeWit2014a}
{Dudok de Wit}, T., Bruinsma, S., \& Shibasaki, K. 2014, Journal of Space
  Weather and Space Climate, 4, A06

\bibitem[{{Dudok de Wit} {et~al.}(2009){Dudok de Wit}, Kretzschmar, Lilensten,
  \& Woods}]{DudokdeWit2009}
{Dudok de Wit}, T., Kretzschmar, M., Lilensten, J., \& Woods, T. 2009,
  Geophysical Research Letters, 36, L10107

\bibitem[{Dulk(1985)}]{Dulk1985}
Dulk, G.~A. 1985, Annual Review, Astronomy \&Astrophysics, 23, 169

\bibitem[{Felli {et~al.}(1981)Felli, Lang, \& Willson}]{Felli1981}
Felli, M., Lang, K.~R., \& Willson, R.~F. 1981, The Astrophysical Journal, 247,
  325

\bibitem[{Foukal(1998)}]{Foukal1998}
Foukal, P. 1998, Geophysical Research Letters, 25, 2909

\bibitem[{Fr\"{o}hlich(2009)}]{Frohlich2009}
Fr\"{o}hlich, C. 2009, Astronomy \& Astrophysics, 501, L27

\bibitem[{Furst {et~al.}(1979)Furst, Hirth, \& Lantos}]{Furst1979}
Furst, E., Hirth, W., \& Lantos, P. 1979, Solar Physics, 63, 257

\bibitem[{Gary(1996)}]{Gary1996}
Gary, D.~E. 1996, in Astronomical Society of the Pacific Conference Series, ed.
  A.~R. Taylor \& J.~M. Paredes, Vol.~93, San Francisco, CA, 387--396

\bibitem[{{Golub} \& {Pasachoff}(2010)}]{Golub2010}
{Golub}, L., \& {Pasachoff}, J.~M. 2010, The Solar Corona, 2nd edn. (Cambridge
  University Press)

\bibitem[{Gopalswamy {et~al.}(1991)Gopalswamy, White, \&
  Kundu}]{Gopalswamy1991}
Gopalswamy, N., White, S.~M., \& Kundu, M.~R. 1991, The Astrophysical Journal,
  379, 366

\bibitem[{Greenstein \& Minkowski(1953)}]{Greenstein1953}
Greenstein, J.~L., \& Minkowski, R. 1953, The Astrophysical Journal, 118, 1

\bibitem[{Hannah \& Kontar(2012)}]{Hannah2012}
Hannah, I.~G., \& Kontar, E.~P. 2012, Astronomy \& Astrophysics, 539, 146

\bibitem[{Henney {et~al.}(2012)Henney, Toussaint, White, \& Arge}]{Henney2012a}
Henney, C.~J., Toussaint, W.~a., White, S.~M., \& Arge, C.~N. 2012, Space
  Weather: The International Journal of Research and Applications, 10, S02011

\bibitem[{Jacchia(1971)}]{Jacchia1971}
Jacchia, L.~G. 1971, {Revised Static Models of the Thermosphere and Exosphere
  with Empirical Temperature Profiles}, Tech. rep., Smithsonian Astrophysical
  Observatory

\bibitem[{Johnson(2011)}]{Johnson2011}
Johnson, R.~W. 2011, Astrophysics and Space Science, 332, 73

\bibitem[{Kundu(1959)}]{Kundu1959}
Kundu, M.~R. 1959, International Astronomical Union, 9, 222

\bibitem[{Kundu(1965)}]{Kundu1965}
---. 1965, Solar Radio Astronomy (Interscience Publishers)

\bibitem[{Landi \& {Chiuderi Drago}(2003)}]{Landi2003}
Landi, E., \& {Chiuderi Drago}, F. 2003, The Astrophysical Journal, 589, 1054

\bibitem[{Landi \& {Chiuderi Drago}(2008)}]{Landi2008}
---. 2008, The Astrophysical Journal, 675, 1629

\bibitem[{Landi {et~al.}(2013)Landi, Young, Dere, {Del Zanna}, \&
  Mason}]{Landi2013}
Landi, E., Young, P.~R., Dere, K.~P., {Del Zanna}, G., \& Mason, H.~E. 2013,
  The Astrophysical Journal, 763, 86

\bibitem[{Lean \& Brueckner(1989)}]{Lean1989}
Lean, J.~L., \& Brueckner, G.~E. 1989, The Astrophysical Journal, 337, 568

\bibitem[{Lemen {et~al.}(2012)Lemen, Boerner, Edwards, Rausch, Schrijver,
  Shine, Shing, Stern, Tarbell, Title, Wolfson, Soufli, Spiller, Gullikson,
  McKenzie, Windt, Golub, Podgorski, Testa, Weber, Akin, Chou, Drake, Duncan,
  Friedlaender, Heyman, Hurlburt, Katz, Kushner, Levay, Lindgren, Mathur,
  McFeaters, Mitchell, Rehse, Springer, Wuelser, Yanari, Bookbinder, Cheimets,
  Caldwell, Deluca, Gates, Park, Bush, Scherrer, Gummin, Smith, Auker, Jerram,
  Pool, Beardsley, Clapp, Lang, \& Waltham}]{Lemen2012}
Lemen, J.~R., Boerner, P.~F., Edwards, C.~G., {et~al.} 2012, Sol. Phys., 275,
  17

\bibitem[{Martyn(1948)}]{Martyn1948}
Martyn, D.~F. 1948, Proceedings of the Royal Society Series A Mathematical and
  Physical Sciences, 193, 44

\bibitem[{Maruyama(2010)}]{Maruyama2010}
Maruyama, T. 2010, Journal of Geophysical Research, 115, A04306

\bibitem[{Maruyama(2011)}]{Maruyama2011}
---. 2011, Journal of Geophysical Research, 116, A08303

\bibitem[{Meyer(1985)}]{Meyer1985}
Meyer, J.-P. 1985, The Astrophysical Journal Supplement Series, 57, 173

\bibitem[{Nakajima {et~al.}(1985)Nakajima, Sekiguchi, Sawa, Kai, Kawashima,
  Kosugi, Shibuya, Shinohara, \& Shiomi}]{Nakajima1985}
Nakajima, H., Sekiguchi, H., Sawa, M., {et~al.} 1985, Publications of the
  Astronomical Society of Japan, 37, 163

\bibitem[{{Nakajima} {et~al.}(1994){Nakajima}, {Nishio}, {Enome}, {Shibasaki},
  {Takano}, {Hanaoka}, {Torii}, {Sekiguchi}, {Bushimata}, {Kawashima},
  {Shinohara}, {Irimajiri}, {Koshiishi}, {Kosugi}, {Shiomi}, {Sawa}, \&
  {Kai}}]{Nakajima_abs1994}
{Nakajima}, H., {Nishio}, M., {Enome}, S., {et~al.} 1994, IEEE Proceedings, 82,
  705

\bibitem[{Parker {et~al.}(1998)Parker, Ulrich, \& Pap}]{Parker1998}
Parker, D.~G., Ulrich, R.~K., \& Pap, J.~M. 1998, Solar Physics, 177, 229

\bibitem[{Pesnell {et~al.}(2011)Pesnell, Thompson, \& Chamberlin}]{Pesnell2011}
Pesnell, W.~D., Thompson, B.~J., \& Chamberlin, P.~C. 2011, Solar Physics, 275,
  3

\bibitem[{Piddington \& Minnett(1951)}]{Piddington1951}
Piddington, J.~H., \& Minnett, H.~C. 1951, The Commonwealth Scientific and
  Industrial Research Organization Australia, 4, 131

\bibitem[{Plowman {et~al.}(2013)Plowman, Kankelborg, \& Martens}]{Plowman2013b}
Plowman, J., Kankelborg, C., \& Martens, P. 2013, The Astrophysical Journal,
  771, 2

\bibitem[{Poduval {et~al.}(2013)Poduval, DeForest, Schmelz, \&
  Pathak}]{Poduval2013a}
Poduval, B., DeForest, C.~E., Schmelz, J.~T., \& Pathak, S. 2013, The
  Astrophysical Journal, 765, 144

\bibitem[{Reames(1999)}]{Reames1999}
Reames, D.~V. 1999, The Astrophysical Journal, 518, 473

\bibitem[{Saint-Hilaire {et~al.}(2012)Saint-Hilaire, Hurford, Keating, Bower,
  \& Gutierrez-Kraybill}]{Saint-Hilaire2012}
Saint-Hilaire, P., Hurford, G.~J., Keating, G., Bower, G.~C., \&
  Gutierrez-Kraybill, C. 2012, Solar Physics, 277, 431

\bibitem[{Scherrer {et~al.}(2011)Scherrer, Schou, Bush, Kosovichev, Bogart,
  Hoeksema, Liu, Duvall, Zhao, Title, Schrijver, Tarbell, \&
  Tomczyk}]{Scherrer2011}
Scherrer, P.~H., Schou, J., Bush, R.~I., {et~al.} 2011, Solar Physics, 275, 207

\bibitem[{Schmahl \& Kundu(1995)}]{Schmahl1995}
Schmahl, E.~J., \& Kundu, M.~R. 1995, Journal of Geophysical Research, 100,
  19851

\bibitem[{Schmahl \& Kundu(1998)}]{Schmahl1998}
Schmahl, E.~J., \& Kundu, M.~R. 1998, in Astronomy Society of the Pacific
  Conference Series, ed. K.~S. Balasubramaniam, J.~Harvey, \& D.~Rabin, Vol.
  140, San Francisco, CA, 387--399

\bibitem[{Schou {et~al.}(2011)Schou, Scherrer, Bush, Wachter, Couvidat,
  Rabello-Soares, Bogart, Hoeksema, Liu, Duvall, Akin, Allard, Miles, Rairden,
  Shine, Tarbell, Title, Wolfson, Elmore, Norton, \& Tomczyk}]{Schou2011a}
Schou, J., Scherrer, P.~H., Bush, R.~I., {et~al.} 2011, Solar Physics, 275, 229

\bibitem[{Selhorst {et~al.}(2014)Selhorst, Costa, {Gim\'{e}nez de Castro},
  Valio, Pacini, \& Shibasaki}]{Selhorst2014}
Selhorst, C.~L., Costa, J. E.~R., {Gim\'{e}nez de Castro}, C.~G., {et~al.}
  2014, The Astrophysical Journal, 790, 134

\bibitem[{Svalgaard \& Hudson(2010)}]{Svalgaard2010a}
Svalgaard, L., \& Hudson, H.~S. 2010, Astronomy Society of the Pacific
  Conference Series, 428, 325

\bibitem[{Swarup {et~al.}(1963)Swarup, Kakinuma, Covington, Harvey, Mullaly, \&
  Rome}]{Swarup1963}
Swarup, G., Kakinuma, T., Covington, A.~E., {et~al.} 1963, The Astrophysical
  Journal, 137, 1251

\bibitem[{Tapping(1987)}]{Tapping1987}
Tapping, K.~F. 1987, Journal of Geophysical Research, 92, 829

\bibitem[{Tapping(2013)}]{Tapping2013}
---. 2013, Space Weather, 11, 394

\bibitem[{Tapping {et~al.}(2007)Tapping, Boteler, Charbonneau, Crouch, Manson,
  \& Paquette}]{Tapping2007}
Tapping, K.~F., Boteler, D., Charbonneau, P., {et~al.} 2007, Solar Physics,
  246, 309

\bibitem[{Tapping {et~al.}(2003)Tapping, Cameron, \& Willis}]{Tapping2003a}
Tapping, K.~F., Cameron, H.~T., \& Willis, A.~G. 2003, Solar Physics, 215, 357

\bibitem[{Tapping \& Charrois(1994)}]{Tapping1994}
Tapping, K.~F., \& Charrois, D.~P. 1994, Solar Physics, 150, 305

\bibitem[{Tapping \& DeTracey(1990)}]{Tapping1990}
Tapping, K.~F., \& DeTracey, B. 1990, Solar Physics, 127, 321

\bibitem[{Tapping \& Vald\'{e}s(2011)}]{Tapping2011}
Tapping, K.~F., \& Vald\'{e}s, J.~J. 2011, Solar Physics, 272, 337

\bibitem[{Tobiska {et~al.}(2008)Tobiska, Bouwer, \& Bowman}]{Tobiska2008}
Tobiska, W.~K., Bouwer, S.~D., \& Bowman, B.~R. 2008, Journal of Atmospheric
  and Solar-Terrestrial Physics, 70, 803

\bibitem[{Vats {et~al.}(1998)Vats, Deshpande, Shah, \& Mehta}]{Vats1998}
Vats, H.~O., Deshpande, M.~R., Shah, C.~R., \& Mehta, M. 1998, Solar Physics,
  181, 351

\bibitem[{White(1999)}]{White1999}
White, S.~M. 1999, Solar Physics, 190, 309

\bibitem[{White \& Kundu(1997)}]{White1997}
White, S.~M., \& Kundu, M.~R. 1997, Solar Physics, 174, 31

\bibitem[{White {et~al.}(1992)White, Kundu, \& Gopalswamy}]{White1992}
White, S.~M., Kundu, M.~R., \& Gopalswamy, N. 1992, The Astrophysical Journal,
  78, 599

\bibitem[{White {et~al.}(2000)White, Thomas, Brosius, \& Kundu}]{White2000a}
White, S.~M., Thomas, R.~J., Brosius, J., \& Kundu, M.~R. 2000, The
  Astrophysical Journal, 534, 203

\bibitem[{Wild {et~al.}(1963)Wild, Smerd, \& Weiss}]{Wild1963}
Wild, J.~P., Smerd, S.~F., \& Weiss, A.~A. 1963, Annual Review, Astronomy
  \&Astrophysics, 1, 291

\bibitem[{Wilson {et~al.}(1987)Wilson, Rabin, \& Moore}]{Wilson1987}
Wilson, R.~M., Rabin, D., \& Moore, R.~L. 1987, Solar Physics, 111, 279

\bibitem[{Zanna {et~al.}(2011)Zanna, Dwyer, \& Mason}]{Zanna2011}
Zanna, G.~D., Dwyer, B.~O., \& Mason, H.~E. 2011, Astronomy \& Astrophysics,
  535, 1

\bibitem[{Zhang {et~al.}(2001)Zhang, Kundu, White, Dere, \&
  Newmark}]{Zhang2001}
Zhang, J., Kundu, M.~R., White, S.~M., Dere, K.~P., \& Newmark, J.~S. 2001, The
  Astrophysical Journal, 561, 396

\bibitem[{Zhang {et~al.}(1998)Zhang, White, \& Kundu}]{Zhang1998}
Zhang, J., White, S.~M., \& Kundu, M.~R. 1998, The Astrophysical Journal, 504,
  L127

\bibitem[{Zirin {et~al.}(1991)Zirin, Baumert, \& Hurford}]{Zirin1991}
Zirin, H., Baumert, B.~M., \& Hurford, G.~J. 1991, The Astrophysical Journal,
  370, 779

\end{thebibliography}

\end{document}